\documentclass[prb,onecolumn,nobibnotes,groupedaddress]{revtex4}

\usepackage{enumerate}
\usepackage{color}
\usepackage{amssymb}
\usepackage{amsmath}
\usepackage{graphicx}
\usepackage{upgreek}
\usepackage{mathtools}
\usepackage{amssymb}
\usepackage{algpseudocode}
\usepackage{algorithm}
\usepackage{braket}
\usepackage[hidelinks]{hyperref}

\DeclareMathOperator*{\argminA}{arg\,min}

\begin{document}

\title{Toward the ``platinum standard" of quantum chemistry on quantum computers: perturbative quadruple corrections in unitary coupled cluster theory}

\thanks{This manuscript has been authored by UT-Battelle, LLC, under Contract No.~DE-AC0500OR22725 with the U.S.~Department of Energy. The United States Government retains and the publisher, by accepting the article for publication, acknowledges that the United States Government retains a non-exclusive, paid-up, irrevocable, world-wide license to publish or reproduce the published form of this manuscript, or allow others to do so, for the United States Government purposes. The Department of Energy will provide public access to these results of federally sponsored research in accordance with the DOE Public Access Plan.}





\begin{abstract}
We propose a non-iterative, \textit{post hoc} correction to the unitary coupled cluster theory with single, double, and triple excitations (UCCSDT) ansatz, which considers the leading-order effects of neglected quadruple excitations. We present two ways to derive this quadruples correction to UCCSDT, henceforth referred to as [Q-6], which leads to an improvement in the correlation energy shown to be correct through sixth-order in many-body perturbation theory (MBPT). A comparison between the UCC-based [Q-6] correction proposed in this work and analogous, ``platinum" standard quadruples corrections proposed in conventional coupled cluster (CC) theory recognizes that [Q-6] is distinct from prior corrections since it is constructed entirely from internally connected components. Although Trotterized (t) and full operator variants of UCCSDT exhibit errors in scans of small molecule potential energy surfaces (PESs) that routinely exceed 1.6 mH, we find that t/UCCSDT[Q-6] is nevertheless able to achieve chemical accuracy as measured by the mean-unsigned error (MUE).  

\end{abstract}

\author{Zachary W. Windom$^{1}$, Luke Bertels$^{1}$, Daniel Claudino$^1$\footnote{\href{mailto:claudinodc@ornl.gov}{claudinodc@ornl.gov}}, and Rodney J. Bartlett$^2$}
\affiliation{$^1$Quantum Information Science Section,\ Oak\ Ridge\ National\ Laboratory,\ Oak\ Ridge,\ TN,\ 37831,\ USA\\
$^2$Quantum Theory Project, University of Florida, Gainesville, FL, 32611, USA }
\maketitle

\section{Introduction}

The \textit{ab initio} prediction of molecular and materials properties intimately depends upon an accurate accounting of a system's electronic structure.\cite{bartlett1989coupled} Unfortunately, the full configuration interaction (FCI), which represents the exact solution to the electronic Schr\"{o}dinger equation (SE) within a single-particle basis set, scales exponentially with respect to system size, and is therefore intractable. This has led to the development of methods that mitigate the underlying calculation cost. To this end, approximations based on low-rank many-body perturbation theory (MBPT) and coupled cluster (CC) theory have been shown to excel at capturing the instantaneous electron interactions associated with ``dynamic" correlations found at equilibrium geometries,\cite{bartlett1981many,bartlett1989coupled} whereas multi-reference (MR) methods, like MR-CI,\cite{lengsfield1984evaluation,woywod1994characterization} are designed to specifically cater to the ``static" correlations that dominate when several electronic configurations are necessary for an adequate, zeroth-order trial wavefunction. Regardless, there appears to be ``no such thing as a free lunch" for existing electronic structure methods: typically, a low-rank electronic structure method is only suitable at capturing a particular electron correlation ``type" existing within a subsection of the potential energy surface (PES) 

 In particular, such methods are of limited value when higher-rank excitation effects dominate the wavefunction ansatz and/or extremely accurate results are needed; such situations necessitate inclusion of higher-rank excitation operators. Methods based on CC are advantageous in this regard because they are known to systematically converge toward the FCI limit. This limit is achieved by increasing the maximal rank of the excitation operator, leading to a hierarchy of well-defined approximations that scale polynomially with respect to system size.\cite{bartlett2007coupled,shavitt2009many} However, short of the FCI limit, single-reference CC theory will be sensitive to non-variational catastrophes which arise whenever the overlap between the single-reference Slater determinant and the exact eigenfunction is small.\cite{paldus1984approximate,chan2004state} In such cases, the cluster operator necessarily has to encode comparatively more information to recover from a defective, mean-field starting point.
 
 The importance of including quadruple excitations, in particular, is known to be impactful.\cite{bartlett1978many,bartlett1979quartic,kucharski1989coupled,windom2024ultimate,windoma2024t2} In instances of ``strong'', static correlations, quadruple excitation effects become large and can even anomalously surpass the importance of double electron excitations.\cite{paldus1982cluster,paldus1984degeneracy,piecuch1990coupled} Similarly, high-accuracy model chemistries which seek to predict enthalpies of formation to within 1 kJ mol$^{-1}$ of experimental results depend on estimates of these effects.\cite{tajti2004heat,bomble2006high,martin1999towards,boese2004w3,karton2006w4} However, the steep $\mathcal{O}(N^{10})$ scaling of methods like CCSDTQ can easily become unaffordable. This has led to ``cheaper" estimates of $T_4$ that scale like $\mathcal{O}(N^9)$, such as (Q)\cite{kucharski1989coupled,kucharski1993coupled,bomble2005coupled} and the so-called ``platinum standard of quantum chemistry, (Q$_{\Lambda}$).\cite{kucharski1998sixth,bomble2005coupled} Further reduction in algorithmic scaling can be extracted without appreciable sacrifice to the energy by invoking the factorization theorem of MBPT, leading to the $\mathcal{O}(N^7)$ [Q$_f$] correction in standard CC theory.\cite{kucharski1986fifth,kucharski1998efficient,thorpe2024factorized} The existing repertoire of perturbative quadruple corrections in standard CC theory
 is a testament to the importance in approximating higher-rank excitation effects in certain situations. However, it should be recognized that the underlying CC hierarchy encompassing quadruples excitations remains sensitive to ``non-variational catastrophes" in pathological situations. 

Upon the future introduction of fault-tolerant quantum computers, which naturally cater to the Hermitian analog to CC known as unitary coupled cluster theory (UCC),\cite{bartlett1989alternative,szalay1995alternative} these prospects are expected to change for the following two reasons. First, infinite-order UCC methods obey the Rayleigh-Ritz variational condition, which prevents the ``non-variational catastrophes'' that plague low-rank CC theories in pathological situations. Second, the UCC ansatz can - in principle - be efficiently represented and prepared using a parameterized quantum circuit\cite{anand2022quantum} which is unlike the corresponding classical UCC algorithm that scales exponentially with respect to system size.\cite{bartlett1989alternative} Similar to standard CC theory, a hierarchy of UCC approximations can also be designed\cite{bartlett1989alternative,kutzelnigg1983quantum,liu2018unitary,hodecker2020third,liu2021unitary,liu2022quadratic} to converge toward FCI via systematic inclusion of all higher-rank excitation operators that modulate up to $n$-fold excitations for an $n$ electron system.

To work within the current technological constraints and minimize the number of (expensive) entangling operations--e.g, CNOT gates-- low-rank (UCC) ansatze are favored.\cite{li2025quantum} However, this necessarily means neglecting higher-rank excitations that may be important in maximizing agreement with the FCI. One way to circumnavigate this to some extent is to use adaptive ansatz, which iteratively select operators from a pool that satisfy a predefined optimization criteria.\cite{grimsley2019adaptive,halder2024noise} Although this leads to a ``minimal", resource-efficient ansatz that can converge toward the FCI, there are trade-offs -- namely, the number of measurements can become exceedingly expensive. 

Prior efforts envisioned the constraints of existing quantum hardware by constructing affordable, low-rank UCC ansatz, enabling \textit{post hoc} perturbative energy corrections that consider missing cluster operator excitation effects. This has led to the [4S] and [6S] methods with double excitations (UCCD) that estimate neglected singles excitation effects and which are correct through fourth and sixth-order in MBPT, respectively.\cite{windom2024attractive} Similarly, we recently proposed a fourth-order triples correction to UCCSD, which we call [T],\cite{windom2024new} similar in spirit to the ``gold standard'' of quantum chemistry.\cite{urban1985towards, raghavachari1989fifth,watts1993coupled} In both cases, we found that perturbatively correcting for neglected cluster operators lead to consistently superior agreement with respect to FCI as compared to the corresponding baseline method. Also of importance, in a scenario where one has access to a quantum computer, construction of these perturbative corrections are performed solely on a classical computer and completely bypass the need for additional quantum resources for increased accuracy beyond the learning of optimal amplitudes.

The current work extends these developments by deriving the leading, sixth-order correction associated with neglected quadruples excitations effects with respect to the UCCSDT ansatz; henceforth referred to as the [Q-6] correction. Alternatively, it is relevant that the proposed workflow can be seen as a hybrid computing strategy wherein UCCSDT amplitudes are converged on a quantum computer (a numerical simulator in this paper), which are subsequently used to compute the [Q-6] correction on a classical computer. We note that calculating [Q-6] amounts to a post-processing step that scales as $\mathcal{O}(N^9)$ on a classical computer, but requires no additional quantum resources beyond the baseline UCCSDT method. 

This paper is organized as follows: the Theory section (\ref{sec:theory}) proposes two different ways of deriving [Q-6] and discuss its differences with existing quadruples corrections in standard CC theory. The Computational Details (\ref{sec:comp}) reports the software and parameters used in the simulations. In the Results and Discussion \ref{sec:results} section we report and discuss the performance of the [Q-6] correction for a collection of potential energy surfaces (PES) of first-row diatomic molecules. We present closing remarks in the Conclusion (\ref{sec:conclusion}) as well as an outlook on the proposed method.
\section{Theory}
\label{sec:theory}

We concern ourselves with the solution to the time-independent Schr\"{o}dinger equation, having the form 

\begin{equation}\label{eq:SE}
    H_N\ket{\Psi}=\Delta E\ket{\Psi},
\end{equation} which is given in terms of the eigenfunction $\ket{\Psi}$, and the correlation energy $\Delta E$ which in turn is defined as the difference between the "exact" (FCI) and mean-field solutions. Electron correlation is readily determined with respect to the normal-ordered Hamiltonian $H_N$
\begin{equation}
\begin{split}
\label{eq:ham}
H_N&=H-\braket{0|H|0}\\
&=\underbrace{\sum_{p}\epsilon_{pp}\{p^{\dagger}p\}}_{\text{$H_0$}}+ \underbrace{\frac{1}{4}\sum_{pqrs}\braket{pq||rs}\{p^{\dagger}q^{\dagger}sr\}}_{\text{$V$}}, \\
& = H_0 + V.
 \end{split} 
\end{equation}

For the overwhelming majority of examples in quantum chemistry, solving for the exact $\ket{\Psi}$ of Equation \ref{eq:SE} is algorithmically unfeasible, which makes the development of accurate and tractable ansatze paramount. In this regard, the current work focuses on UCC theory, which parametrizes the wavefunction as an exponential 
\begin{equation}\label{eq:untrunatedUCCansatz}
  \ket{\Psi}_{UCC}\equiv e^{\tau}\ket{0},
\end{equation} where an anti-Hermitian cluster operator $\tau$ acts on mean-field reference determinant $\ket{0}$, which is the canonical Hartree-Fock solution throughout this work. In the limit where $\tau$ accounts for all possible de-excitations/excitations in an $N$-electron system, $\tau = \sum^N_n \tau_n$, and
\begin{equation}
\begin{split}
    \tau_n &= T_n - T_n^{\dagger} \\
    & = \frac{1}{(n!)^2}\sum_{ab\cdots ij\cdots} \big(t_{ij\cdots}^{ab\cdots}\{a^{\dagger}b^{\dagger}ij\}-t_{ab\cdots}^{ij\cdots}\{i^{\dagger}j^{\dagger}ab\}\big)
\end{split}
\end{equation} where $T_n$ and its adjoint are the standard cluster operators. Fortunately, CC theory and its various flavors are known to recover a majority of correlation effects, even in its low-rank formulations, as compared to FCI. However, in some circumstances low-rank CC may not be immediately capable of achieving a desired threshold for accuracy. A traditional response to this issue has been to cheaply estimate the correlation effects of neglected cluster operators, which is an attempt at increasing the fidelity with respect to FCI.\cite{urban1985towards,raghavachari1989fifth,bartlett1990non,stanton1997ccsd,kucharski1989coupled,bomble2005coupled}

Our objective in this work is to develop a rigorous way to cheaply account for neglected quadruples excitations to the UCCSDT ansatz by approximating the leading-order effects of $\tau_4$ using tenets of perturbation theory. In the following two subsections, we outline two separate ways in which to derive the [Q-6] perturbative correction to UCCSDT. 

\subsection{L\"{o}wdin partitioning approach}

A formal solution to the SE equation can be expressed in terms of the Hilbert space spanned by all $N$-electron Slater determinants, which consist of the HF determinant in addition to all determinants representative of single, double, triple, up to $N$-tuple excitations out of the HF reference. This complete space, $\ket{\textbf{h}}$, can be partitioned into two subspaces: an ``important", but easy-to-solve portion, called $\ket{\textbf{p}}$, and a less-important part called $\ket{\textbf{q}}$, such that $\ket{\textbf{h}} = \ket{\textbf{p}} \bigoplus \ket{\textbf{q}}$. As our derivations are with respect to UCCSDT, we explicitly define $\ket{\textbf{p}} = \ket{0}\bigoplus \ket{\textbf{s}}\bigoplus \ket{\textbf{d}}\bigoplus \ket{\textbf{t}}$, where $\ket{0}$ is the HF determinant, and bolded letters \textbf{s}, \textbf{d}, and \textbf{t} refer to the set of singly, doubly, and triply excited determinants. 

Using the matrix partitioning approach originally proposed by L\"{o}wdin,\cite{lowdin1962studies} we can represent the exact solution to the SE in terms of the components of $\ket{\textbf{h}} $, and a unitarily-transformed Hamiltonian $\bar{H}$ defined with respect to the UCCSDT wavefunction
\begin{equation}
  \bar{H} = e^{-\tau_1-\tau_2-\tau_3}H_Ne^{\tau_1+\tau_2+\tau_3}\equiv\bigg( H_Ne^{\tau_1 + \tau_2 + \tau_3}\bigg)_C.
\end{equation} 
Note that our definition of $|\textbf{p}\rangle$ can be used to represent the UCCSDT equations:

\begin{subequations}
\begin{align}
\braket{0|\bar{H}|0}=E _\text{UCCSDT} \label{eq:UCCSDThbareiga}\\
\braket{\textbf{p}|\bar{H}|0}=0 \label{eq:UCCSDThbareigb}
\end{align}
\end{subequations}

 where Equation \ref{eq:UCCSDThbareiga} represents the UCCSDT correlation energy in terms of an eigenvalue problem having the HF determinant as a solution, and Equation \ref{eq:UCCSDThbareigb} is the UCCSDT residual equations which determine the cluster amplitudes.

After inserting this definition along with the resolution of the identity into the SE, followed by projecting the result onto $\ket{\textbf{p}}$ and $\ket{\textbf{q}}$,  we see that the SE can be re-expressed as
\begin{equation}\label{eq:smallHbar}
  \begin{pmatrix}
    \bar{H}_{\textbf{p}\textbf{p}} & \bar{H}_{\textbf{p}\textbf{q}} \\
    \bar{H}_{\textbf{q}\textbf{p}} & \bar{H}_{\textbf{q}\textbf{q}}
  \end{pmatrix}
  \begin{pmatrix}
    C_{\textbf{p}}\\
    C_{\textbf{q}}
  \end{pmatrix} = E
    \begin{pmatrix}
    C_{\textbf{p}}\\
    C_{\textbf{q}}
  \end{pmatrix},
\end{equation}
where $C_{\textbf{p}} = \ket{\textbf{p}}\braket{\textbf{p}|\Psi}$ and $C_{\textbf{q}} = \ket{\textbf{q}}\braket{\textbf{q}|\Psi}$ are projections of the exact eigenfunction $\ket{\Psi}$ onto the $\textbf{p}$ and $\textbf{q}$, respectively. We can expand the above Equation \ref{eq:smallHbar} using the definition of $\ket{\textbf{p}}$ and $\ket{\textbf{q}}$ to rewrite the effective Hamiltonian $\bar{H}$  in a more transparent way:

\begin{equation}\label{eq:explicitHbar}
\bar{H}= \begin{pmatrix}
\bar{H}_{00} & \bar{H}_{0\textbf{p}} & \bar{H}_{0\textbf{q}}\\
\bar{H}_{\textbf{p}0} & \bar{H}_{\textbf{p}\textbf{p}} & \bar{H}_{\textbf{p}\textbf{q}}\\
\bar{H}_{\textbf{q}0} & \bar{H}_{\textbf{q}\textbf{p}} & \bar{H}_{\textbf{q}\textbf{q}}\\
\end{pmatrix} \equiv \begin{pmatrix}
E _\text{UCCSDT} & 0 & \bar{H}_{0\textbf{q}}\\
0 & \bar{H}_{\textbf{p}\textbf{p}} & \bar{H}_{\textbf{p}\textbf{q}}\\
\bar{H}_{\textbf{q}0} & \bar{H}_{\textbf{q}\textbf{p}} & \bar{H}_{\textbf{q}\textbf{q}}\\
\end{pmatrix},
\end{equation} 
where $\bar{H}_{00}$, $\bar{H}_{\textbf{p}0}$, and $\bar{H}_{0\textbf{p}}$ are the UCCSDT equations found in Equation \ref{eq:UCCSDThbareigb}. From Equation \ref{eq:smallHbar}, it is straightforward to show 
$C_{\textbf{q}}=\bigg( E - \bar{H}_{\textbf{q}\textbf{q}}\bigg)^{-1} \bar{H}_{\textbf{q}\textbf{p}} C_{\textbf{p}}$ which is inserted into the remaining linear equation to form an ``effective" eigenvalue problem
\begin{equation}\label{eq:Heff}
\begin{split}
    \bar{H}_{eff}C_{\textbf{p}} &\equiv \bigg(\bar{H}_{\textbf{p}\textbf{p}} + \bar{H}_{\textbf{p}\textbf{q}}(E - \bar{H}_{\textbf{q}\textbf{q}} )^{-1} \bar{H}_{\textbf{q}\textbf{p}}\bigg)C_{\textbf{p}}=EC_{\textbf{p}}\\
    &\Rightarrow C_{\textbf{p}}^{\dagger}\bar{H}_{\textbf{p}\textbf{p}}C_{\textbf{p}}+ C_{\textbf{p}}^{\dagger}\bar{H}_{\textbf{p}\textbf{q}}(E - \bar{H}_{\textbf{q}\textbf{q}} )^{-1} \bar{H}_{\textbf{q}\textbf{p}}C_{\textbf{p}} = E C_{\textbf{p}}^{\dagger}C_{\textbf{p}}, \\
\end{split}
\end{equation} 
with $\bigg(E-E _\text{UCCSDT}\bigg) \equiv \Delta E $. Note that the benefit we extract from following this protocol is that the eigenvalue problem for $\bar{H}$, whose matrix representation originally had a basis spanning the Hilbert space, is now equivalently represented in terms of an ``effective" operator that is of the same rank as the much smaller $\ket{\textbf{p}}$. Nevertheless, we are still limited by inverting a matrix that is of the same rank as the $\ket{\textbf{q}}$. However, we can partition $\bar{H}$ into a zeroth-order and perturbative component, at which point many-body perturbation theory can be used to expand Equation \ref{eq:Heff} on an order-by-order basis. 

Our expansion for $\bar{H}$ is done with respect to the Moller-Plesset fluctuation potential, where we count ``orders" based on an operators' initial contribution to electron correlation with respect to a canonical HF reference. In that case, $V$ and $\tau_2$ are first-order, $\tau_1$ and $\tau_3$ are second-order, $\tau_4$ is third-order, and so on. With this in mind, our expansion in $\bar{H}$ appears as

\begin{equation}\label{eq:newhbardef}
\begin{split}
\bar{H}^{[0]} &= H_0\\
     \bar{H}^{[1]}& = V \\
\bar{H}^{[2]} &= [V,\tau_2] \\
    \bar{H}^{[3]} &= [V,\tau_3] + [V,\tau_1] + \frac{1}{2}[[V,\tau_2],\tau_2]] \\
    & \vdots \\ 
\end{split}
\end{equation} Note that the above omits nested commutators involving $H_0$, since they cannot project onto the ($T_4$-portion of) $\ket{\textbf{q}}$ at such low-orders in $\bar{H}$. Furthermore, such terms do not contribute to the energy expression in finite-order UCC theories.\cite{szalay1995alternative} We similarly represent $C_{\textbf{p}}$ in an many-body expansion such that
\begin{equation}\label{eq:COp}
  C_{\textbf{p}} = 1 + C_{\textbf{p}}^{[4]} + C_{\textbf{p}}^{[5]} + \cdots
\end{equation} where we recognize that corrections to the UCCSDT ansatz associated with low-order approximations to $\tau_4$, projected onto the $|\mathbf{p}\rangle$, initially arise at fourth-order in MBPT. 

Using Equation \ref{eq:newhbardef}, the underlying resolvent operator can be expressed recursively as 
\begin{equation}
\label{eq:reoslvent}
  R(E) = (E_0-QH_0Q)^{-1} + (E_0-QH_0Q)^{-1}\bigg( \bar{H}^{[1]} + \bar{H}^{[2]} + \bar{H}^{[3]}+ \bar{H}^{[4]}+ \cdots\bigg)R(E)
\end{equation} 
Insertion of Equations \ref{eq:newhbardef}, \ref{eq:COp}, and \ref{eq:reoslvent}  back into Equation \ref{eq:Heff} leads to an expression that determines $\tau_4$ contributions to the energy starting at sixth-order which we refer to as [Q-6]:
\begin{equation}\label{eq:sqrbrakQ}
\begin{split}
\Delta E^{[Q-6]} = &\braket{0|\bar{H}^{[3]}|\textbf{Q}}\braket{\textbf{Q}|E_0 - \textbf{Q}H_0\textbf{Q}|\textbf{Q}}^{-1}\braket{\textbf{Q}|\bar{H}^{[3]}|0} \\
     =& \braket{0|\bigg(T_3^{\dagger}W_N\bigg)_C\textbf{D}_4\bigg(W_NT_3\bigg)_C|0} + \frac{1}{2}\braket{0|\bigg(T_3^{\dagger}W_N\bigg)_C\textbf{D}_4\bigg(W_NT_2^2\bigg)_C|0}\\
    &+\frac{1}{2}\braket{0|\bigg((T_2^{\dagger})^2W_N\bigg)_C\textbf{D}_4\bigg(W_NT_3\bigg)_C|0}+\frac{1}{4}\braket{0|\bigg((T_2^{\dagger})^2W_N\bigg)_C\textbf{D}_4\bigg(W_NT_2^2\bigg)_C|0}
\end{split}
\end{equation}

\subsection{Sixth-order UCC functional}
Alternatively, we can conceive of a quadruples correction with respect to the sixth-order UCC energy functional, similar to related work studying the expectation-value coupled cluster energy functional.\cite{kucharski1998noniterative,kucharski1998sixth} This route necessitates the cancellation of so-called internally disconnected diagrams which becomes increasingly tedious as the number of terms in the energy functional grow. For brevity, we highlight the pertinent aspects of this approach in the following. 

The strict UCCSDT energy functional  yielding correlation energies which are correct through sixth-order in MBPT can be expressed generically in terms of
\begin{equation}\label{eq:ucc6uccsdt}
   \Delta E _\text{UCCSDT(6)} \equiv \braket{0|e^{\tau_1^{\dagger}+\tau_2^{\dagger}+\tau_3^{\dagger}}H_Ne^{\tau_1+\tau_2+\tau_3}|0}
\end{equation} As written, this deceivingly simple form neglects reference to any of the simplifications that could be invoked, and is therefore an abstract representation for the formal UCCSDT(6) equations that nevertheless is enough for our immediate purposes. A more pertinent idea to recognize is that the UCCSDT(6) energy functional of Equation \ref{eq:ucc6uccsdt}, in principle, yields a series of residual equations that, upon solution, generate a converged set of   $\tau_1$, $\tau_2$, and $\tau_3$ amplitudes. Once these amplitudes are obtained, we then attempt to transcend this approximation by constructing a (presumably) more accurate UCC energy functional that considers the leading-order effects of $\tau_4$ on the sixth-order energy, in addition to  $\Delta E _\text{UCCSDT(6)}$. We generically refer to this more ``complete" energy functional, $\Delta E^{[6]}_{UCC}$, as 
\begin{equation}\label{eq:basicUCC6}
\Delta E^{[6]}_\text{UCC} = \Delta E^{[6]}_\text{UCC}(SDT) + \Delta E^{[6]}_\text{UCC}(Q)
\end{equation} where $\Delta E^{[6]}_{UCC}(SDT)$ is the the portion of Equation  \ref{eq:ucc6uccsdt} that is correct through sixth-order in MBPT and $\Delta E^{[6]}_{UCC}(Q)$ is the remaining part of the functional that depends on quadruples excitation operators
\begin{equation}\label{eq:t4dependence}
    \Delta E^{[6]}_{UCC}(Q) = \braket{0|T_4^{\dagger}f_NT_4 + \bigg( T_4^{\dagger}W_NT_3 + \text{h.c.}\bigg) + \frac{1}{2}\bigg(T_4^{\dagger}W_NT_2^2+ \text{h.c.} \bigg)|0}
\end{equation} 

Clearly, Equation \ref{eq:t4dependence} expresses energy contributions arising from $\tau_4$ via converged $\tau_1$, $\tau_2$, and $\tau_3$ amplitudes. Variation of $\Delta E^{[6]}_\text{UCC}(Q)$ with respect to $\tau_4^{\dagger}$ determines the corresponding set of amplitudes. In other words, this functional's $T_4$ residual equations satisfy $\frac{\partial \Delta E(6)}{\partial T_4^{\dagger}} = 0 $, which leads to 
\begin{equation}\label{eq:t4resideqns}
\begin{split}
   & Q_4 \bigg( f_NT_4 + W_NT_3 + \frac{1}{2}W_NT_2^2 \bigg) = 0  \\
   & \Rightarrow D_4T_4 = W_NT_3 + \frac{1}{2}W_NT_2^2\\
   &\Rightarrow T_4^{[3]} = \frac{1}{D_4}\bigg( W_NT_3 + \frac{1}{2}W_NT_2^2\bigg)
\end{split}
\end{equation} 
where $T_4^{[3]}$ is the leading-order approximation to $T_4$ that is correct through third-order in MBPT. Insertion of the $T_4^{[3]}$ amplitudes back into  the (third term) of Equation \ref{eq:basicUCC6} nullifies this term's contribution to the final energy. Consequently, we only consider the second term of Equation \ref{eq:basicUCC6}, and only contributions  which arise at sixth-order in MBPT therein. The final form of the $\Delta E(6)$ energy functional appears as 
\begin{equation}\label{eq:simplifiedUCC6}
  \Delta E(6) = \Delta E _\text{UCCSDT(6)} + \braket{0|T_3^{\dagger}W_NT_4 + \frac{1}{2}\bigg((T_2^{\dagger})^2W_N\bigg)_CT_4|0}
\end{equation} 

Inserting the definition for  $T_4^{[3]}$ found in Equation \ref{eq:t4resideqns} into Equation \ref{eq:simplifiedUCC6} leads to the same definition for [Q-6] as introduced in  Equation \ref{eq:sqrbrakQ}.

 \subsection{Differences between UCC and CC quadruples corrections}
\begin{figure}[ht!]
\centering
\includegraphics[width=\columnwidth]{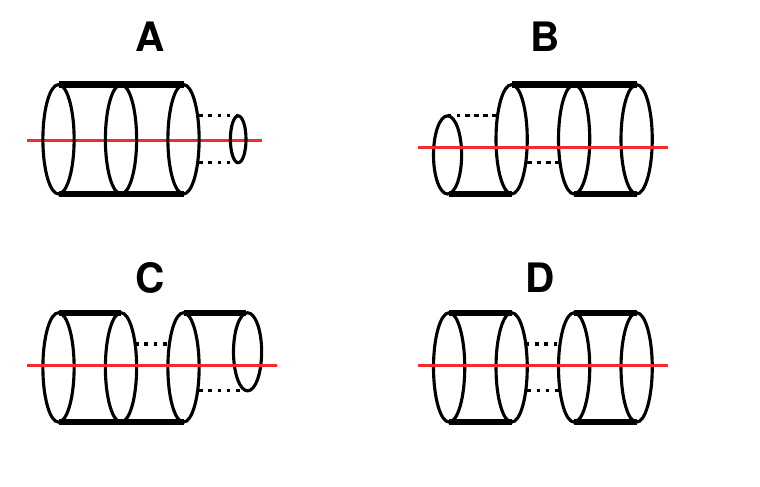}  

\caption{Illustration of the skeleton diagrams defining the [Q] correction in UCC theory. Horizontal red lines indicate the 8-index Fock denominator $\epsilon_{ijkl}^{abcd} = (\epsilon_i+\epsilon_j+\epsilon_k+\epsilon_l-\epsilon_a-\epsilon_b-\epsilon_c-\epsilon_d)^{-1}$, bold black lines indicate cluster operators $T_n$, and dotted lines indicate two-electron integrals $\langle pq||rs \rangle$.}
\label{fig:Qcorr}
\end{figure}

The [Q-6] correction defined in Equation \ref{eq:sqrbrakQ} is translated diagrammatically in Figure \ref{fig:Qcorr}, where we note that diagrams A and B are two of the four diagrams participating in the conventional (Q)/(Q)$_{\Lambda}$ correction, known equivalently as the ``platinum" standard in standard CC theory. However, diagrams C and D are completely unique to UCC and appear as completely connected counterparts to the remaining two diagrams in (Q)/(Q)$_{\Lambda}$. The (Q)/(Q)$_{\Lambda}$ approaches in standard CC theory ``caps'' diagrams C and D of Figure \ref{fig:Qcorr} using $(T_2^{\dagger})^2$ instead of the fully connected term $((T_2^{\dagger})^2W_N)_C$.

 This discrepancy between the UCC [Q-6] and CC (Q)/(Q)$_{\Lambda}$ corrections make intuitive sense, especially when one recognizes that developing finite-order approximations to UCC theory  involves the tedious elimination of terms that are deemed ``internally disconnected". These internally disconnected diagrams lead to unlinked energy expressions, meaning the result would not be size-extensive; although, once the UCC equations are solved to infinite-order these terms are naturally eliminated. With this in mind, terms like $(T_2^{\dagger})^2D_4(W_NT_3)_C + \text{h.c.}$ in standard (Q)/(Q)$_{\Lambda}$ would necessarily contribute a set of disconnected terms in the $T_3$/$T_3^{\dagger}$ residual equations, ultimately leading to unlinked energy diagrams and a subsequent loss in size-extensivity. The UCC formulation for [Q-6] avoids this issue by construction, as can be visualized in the added connectivity of diagrams C and D, and further explains why the analogous internally disconnected diagrams in the standard (Q)/(Q)$_{\Lambda}$ correction cannot participate in the UCC energy functional. 
 
\section{Computational Details}
\label{sec:comp}

All UCC calculations are performed in the XACC software.\cite{mccaskey2020xacc, xacc_chem} We determine $\tau_1$, $\tau_2$, and $\tau_3$ via the Variational Quantum Eigensolver (VQE) using a numerical simulator that relies on PySCF\cite{sun2018pyscf} for molecular integrals. The STO-6G basis set is used throughout this work,\cite{Hehre1969,Feller1996,Pritchard2019} and all examples drop core electrons from the correlation calculation. The resulting set of $\tau$ amplitudes are externally relayed to the pyCC software\cite{UT2} which computes the [Q-6] perturbative correction to UCCSDT. The current pilot implementation for [Q-6] is constructed according to the standard convention\cite{shavitt2009many}
\begin{equation}
  \Delta E(6) = \frac{1}{(4!)^2}\sum_{ijkl,abcd} (t_{ijkl}^{abcd ^{[3]}}\epsilon_{ijkl}^{abcd}) (t_{ijkl}^{abcd^{[3]}})^{*}
\end{equation} where the first term in parenthesis is given by the amplitude found in the last line of Equation \ref{eq:t4resideqns} and the second term in parenthesis is given by the residual equation found in the second line of Equation \ref{eq:t4resideqns}.

As reported in the work of Olsen and Kohn,\cite{kohn2022capabilities} UCCSDT is consistently in very close agreement with FCI for most commonly studied, minimal basis set examples that are conventionally thought to exhibit static correlation and/or multi-reference effects. To emphasize this point, their analysis of twisted ethylene in a minimal basis set -- which might traditionally be considered a 4 electron active space problem -- reports a UCCSDT error below 1 mH for all twist angles. On top of this, perturbative corrections to CC are known to be sensitive to static correlations, although there is some evidence that suggests such corrections built with respect to UCC amplitudes can potentially be more robust when scanning a PES.\cite{windom2024attractive,windom2024new} When considering these issues, we chose to benchmark the [Q-6] correction to UCCSDT on a test set of diatomic molecules whose initial selection was guided by prior literature,\cite{lee2019kohn,windom2022benchmarking,windom2024attractive} and which were subsequently found to require an accounting of higher-order excitation effects beyond the baseline, minimal basis set UCCSDT. Our test set consists of LiF, NF, BO$^-$, N$_2$, and O$_2$, all in their lowest energy, singlet electronic configuration. All calculations are with respect to a restricted Hartree-Fock (RHF) reference determinant.

Individual cluster operators do not, in general, commute with each other (e.g. $[T_i,T_j] \neq 0$). This leads to the following peculiarity of the UCC ansatz when compared to conventional CC theory: the sum of the product of UCC exponentials is not, in general, equal to the product of the sum of UCC exponentials. This fact naturally lends itself to two distinct ``flavors" of UCC ansatz. We refer to the first flavor as the ``full" UCCSDT ansatz, which is defined according to
\begin{equation}
  \label{eq:uccd}
  |\Psi_\text{UCCSDT}\rangle = e^{\sum_{ia}\theta_{i}^{a}(a^\dagger i- \text{h.c.})+\sum_{ijab}\theta_{ij}^{ab}(a^\dagger b^\dagger ij- \text{h.c.})+\sum_{ijkabc}\theta_{ijk}^{abc}(a^\dagger b^\dagger c^\dagger ijk- \text{h.c.})} |0\rangle.
\end{equation}
The second flavor of UCC ansatz is based on trotterization (t) of the above, henceforth referred to as tUCCSDT:
\begin{align}
   |&\Psi_\text{tUCCSDT}\rangle = \prod_{ia}e^{\theta_i^a(a^\dagger i- \text{h.c.})}\prod_{ijab}e^{\theta_{ij}^{ab}(a^{\dagger}b^{\dagger}ji-\text{h.c.})}\prod_{ijkabc}e^{\theta_{ijk}^{abc}(a^{\dagger}b^{\dagger}c^{\dagger}jik-\text{h.c.})}|0\rangle,
  \label{eq:ansatz}
\end{align} In the context of the current work, we explore the effects of adding the [Q-6] correction to both ``flavors" of UCCSDT ansatz. The variational quantum eigensolver (VQE)\cite{Peruzzo2014} is used to obtain the $\tau$ amplitudes, which minimize the expectation value of the Hamiltonian in Equation \ref{eq:ham}
\begin{equation}
   \tau_1^*,\tau_2^*,\tau_3^* = \argminA_{ \tau_1,\tau_2,\tau_3}\braket{\Psi(\tau_1,\tau_2,\tau_3)| H | \Psi( \tau_1,\tau_2,\tau_3)},
\end{equation}
with the set of final, converged $\tau_2^*,\tau_3^*$ being used to construct the [Q-6] correction to the UCCSDT energy, illustrated in Figure \ref{fig:Qcorr}. We note that the operator ordering defining the trotterized UCCSDT ansatz is not necessarily the same in all PES examples studied in this work. To be clear, the adopted operator ordering was predicated on an ``acceptable" [Q-6]  correction. In this context, a particular operator ordering used to construct the [Q-6] correction was deemed ``acceptable" if general agreement was found with the full UCCSDT[Q-6] results.  For LiF and NF, this procedure ultimately resulted in operator orderings that were recommended in prior work.\cite{grimsley2019trotterized}  On the other hand, this default operator ordering was found to yield erroneous [Q-6] corrections for BO$^-$, N$_2$, and O$_2$. Consequently, we adopt an operator ordering that is the ``reverse"  of the  default ordering as this choice was found to yield [Q-6] results that largely coincide with full operator UCCSDT[Q-6]. One way we ascertained this was by decomposing the overall [Q-6] correction into individual, diagrammatic contribution; the final Results and Discussion subsection provides this analysis to some extent. Additional commentary covering our choices for tUCCSDT definition, as well as numerical illustrations highlighting the impact of operator ordering choice  on the [Q-6] correction is relegated to the Supplementary Material.

\section{Results and Discussion}
\label{sec:results}

Table \ref{tab:errors} presents a summary of our findings for the PESs that we studied in terms of the mean-unsigned error (MUE) and the non-parallelity error (NPE). Higher-rank excitation operators are clearly necessary for an accurate representation of the chosen systems' electronic structure, which is evident by the t/UCCSD and t/UCCSDT results. Accounting for only singles and doubles excitation operators leads to -- at minimum -- 5 mH MUE and 7 mH NPE. Explicit inclusion of triples excitation operators leads to t/UCCSDT broadly reducing the error of t/UCCSD by at least half. Still, t/UCCSDT alone is not routinely capable of achieving chemical accuracy with the only notable exception being the MUE for O$_2$. This broadly indicates that a description of higher-rank excitation operators is necessary for these examples, giving us ample opportunity to assess potential benefits in augmenting the t/UCCSDT ansatz with the [Q-6] correction. The following analyzes the contents of Table \ref{tab:errors} in the context of each molecular system. 

\begin{table}[ht!]
    \centering
    \caption{Mean unsigned error (and NPE in parenthesis) with respect to FCI, in mH. Note that the operator orderings are consistently applied for trotterized ansatz. }
    \label{tab:errors}
\begin{tabular}{|c| |c |c |c |c |c|} 
 \hline
Method & LiF & NF & BO$^-$ & N$_2$ & O$_2$ \\ \hline
UCCSD	&11.36 (28.25)&	9.26 (19.00)&	22.70 (51.91)&	6.21 (16.29)&	5.94 (7.48) \\
UCCSDT	&5.01 (14.76)&	2.53 (7.29)	&7.73 (27.35)	&4.97 (16.86)	&1.55 (3.39)\\
UCCSDT[Q-6]	&1.72 (5.95)&	1.74 (5.80)&	0.90 (14.89)&	-0.05 (5.81)	&0.49 (2.84)\\
tUCCSD	&11.04 (27.60)&	9.26 (19.07)&	22.92 (51.54)	&5.61 (13.78)&	5.94 (7.48)\\
tUCCSDT	&4.50 (13.74)&	2.56 (7.44)	&8.35 (28.86)	&4.37 (14.39)	&1.65 (3.56)\\
tUCCSDT[Q-6]	&0.86 (4.09)&	1.88 (6.19)&	1.16 (15.49)&	0.30 (6.83)&	0.83 (2.64)\\ \hline

\end{tabular}

\end{table}

\subsection{LiF}
\begin{figure}[ht!]
 \centering
 \includegraphics[width=\columnwidth]{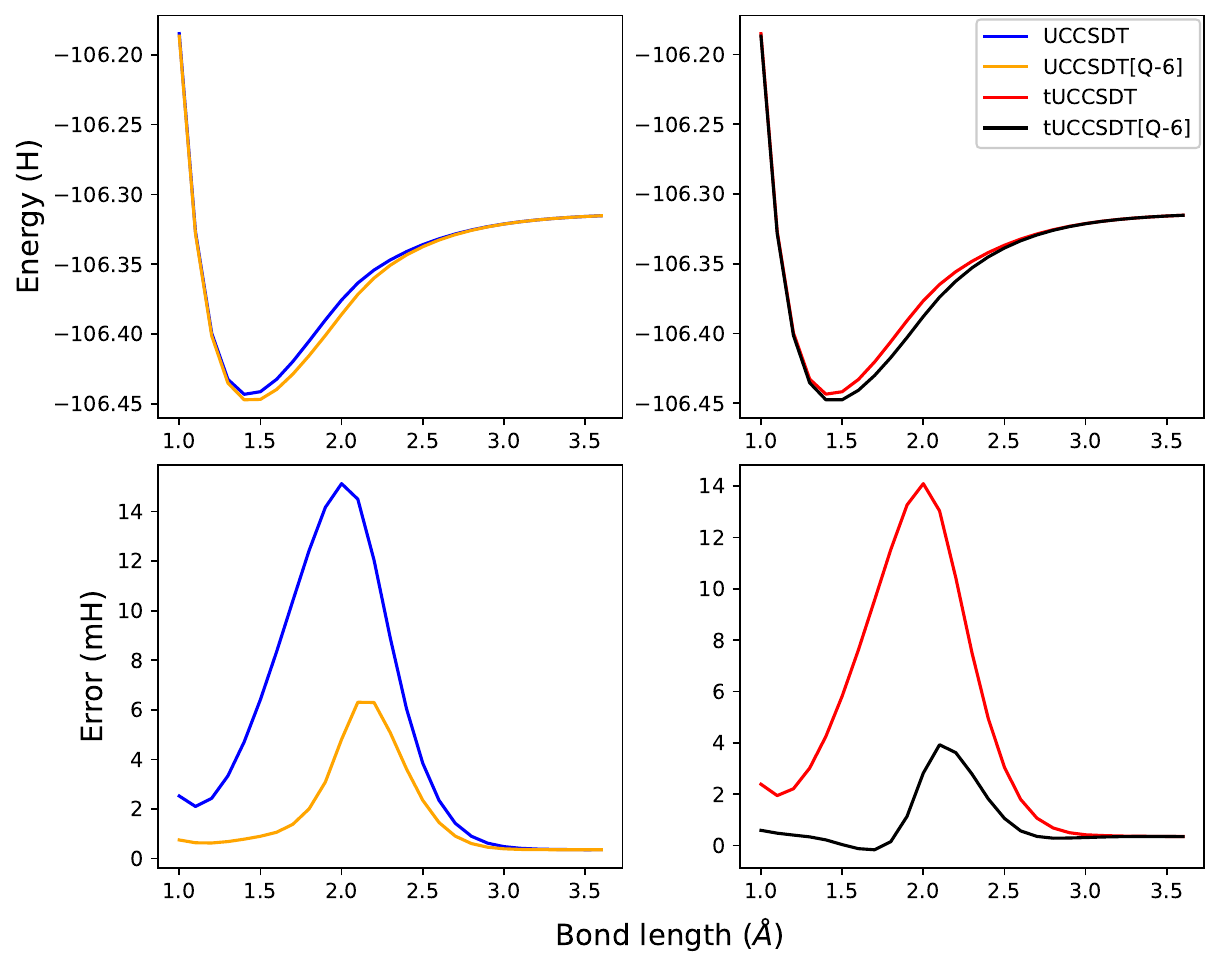}
 \caption{Potential energy surfaces computed with UCCSDT/[Q-6] and tUCCSDT/[Q-6] and corresponding errors with respect to FCI for the dissociation of LiF.}
 \label{fig:LiFpes}
\end{figure}

We begin our analysis with the LiF PES, shown in Figure \ref{fig:LiFpes}. Upon dissociation, there is a near-degeneracy between the ionic and covalent singlet electronic states of LiF which ultimately leads to an avoided crossing amongst these PESs. This has made the LiF PES an attractive target when assessing methods based on the complete-active space (CAS) approach,\cite{casanova2012avoided} since correlated methods based on a single Slater determinant are traditionally thought of as being inadequate. Indeed, our previous work revealed that UCCSD exhibits an anomalously large error with respect to FCI for this example, which performed significantly worse than conventional CCSD.\cite{windom2024ultimate} This is similarly indicated by the results in Table \ref{tab:errors}, highlighting that t/UCCSD exhibits a MUE of 11 mH and a NPE of more than 27 mH. These errors are effectively cut in half by the t/UCCSDT ansatz, which incurs a MUE and NPE that is around 5 and 13 mH, respectively.  However, the explicit inclusion of triples excitation operators is still not enough to achieve the threshold for chemical accuracy. 

The benefits from including the [Q-6] correction are clear in Figure \ref{fig:LiFpes}, where we immediately recognize that t/UCCSDT[Q-6] is superior to baseline t/UCCSDT across the entire LiF PES.  In between 1-1.5 \AA, the [Q-6] correction reduces the UCCSDT error to below 1 mH, while the error between points 1.6-2.2 {\AA} is reduced by more than half. At their worst, UCCSDT is in error with respect to FCI by $\approx$15 mH whereas UCCSDT[Q-6] reduces this value by roughly a third ($\approx$6 mH). These general trends persist even when trotterizing the UCCSDT ansatz, as shown by the tUCCSDT/tUCCSDT[Q-6] results in Figure \ref{fig:LiFpes}; in fact, the success of tUCCSDT[Q-6] seems to be accentuated to some extent. In this case, the [Q-6] correction reduces the error of tUCCSDT to below a mH between 1-1.8 {\AA}. Perhaps more impressive than this, the maximum error of tUCCSDT[Q-6] with respect to FCI is roughly 4 mH which is significantly better than the maximum of $\approx$14 mH found by tUCCSDT. Referring to Table \ref{tab:errors}, we note that the [Q-6] correction reduces the MUE of t/UCCSDT from roughly 5 mH to 1.72 and 0.86 mH, respectively, while the corresponding [Q-6] NPE is a third of the baseline t/UCCSDT results.

\subsection{NF}
\begin{figure}[ht!]
 \centering
 \includegraphics[width=\columnwidth]{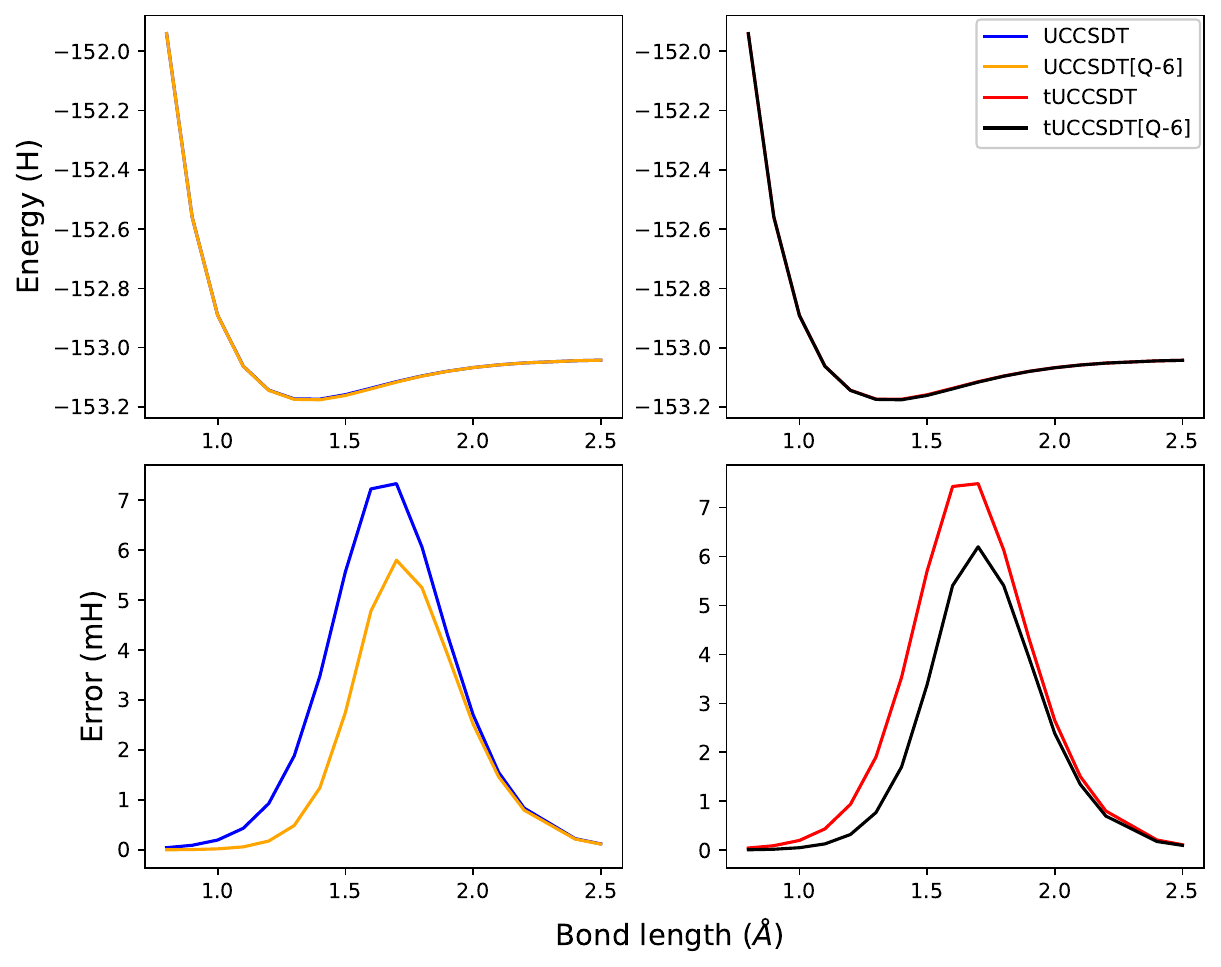}
 \caption{Potential energy surfaces computed with UCCSDT/[Q-6] and tUCCSDT/[Q-6] and corresponding errors with respect to FCI for the dissociation of NF in the lowest energy, singlet ground state.}
 \label{fig:NFpes}
\end{figure}

Figure \ref{fig:NFpes} illustrates the performance of t/UCCSDT and the impact of the [Q-6] correction for the NF molecule. For this example, there are two low-lying, singlet excited states which are in close proximity: a$^1\Delta$ and b$^1\Sigma^+$. These excited states are of interest, since it was found that each has anomalously large dipole moments that counterintuitively point toward $N^-F^+$.\cite{harbison2002electric} Other atypical attributes of NF, such as the excited singlet states having smaller equilibirum bond length than the ground state, suggest a non-standard, complicated electronic structure. Multi-reference methods have previously been employed to study the singlet excited state surfaces.\cite{harbison2002electric,kardahakis2005multireference,su2008valence} It is generally clear from Figure \ref{fig:NFpes} that the single-reference t/UCCSDT method -- bolstered by the [Q-6] correction -- is capable of reproducing the FCI results. 
 
Inspection of Table \ref{tab:errors} emphasizes that t/UCCSDT[Q-6] reduces the MUE to the cusp of chemical accuracy, while reducing the NPE by 1-2 mH. Analyzing these results in more detail, we find that the UCCSDT method already exhibits errors with respect to FCI that are below 1 mH for 0.8-1.2 {\AA} and 2.2-2.5 {\AA} which can be further reduced by half once the [Q-6] correction is added. Similarly, the [Q-6] correction also decreases the UCCSDT error by roughly half between 1.3-1.6 {\AA}. The UCCSDT[Q-6] method exhibits a maximum error of about 5.8 mH, as compared to approximately 7.3 mH maximum error of UCCSDT. These general trends are also followed when the ansatz is trotterized, in which case the maximum error of tUCCSDT becomes closer to 10.5 mH and is brought down to roughly 6.2 mH when adding the [Q-6] correction on top of baseline tUCCSDT.

\subsection{BO$^-$}
\begin{figure}[ht!]
 \centering
 \includegraphics[width=\columnwidth]{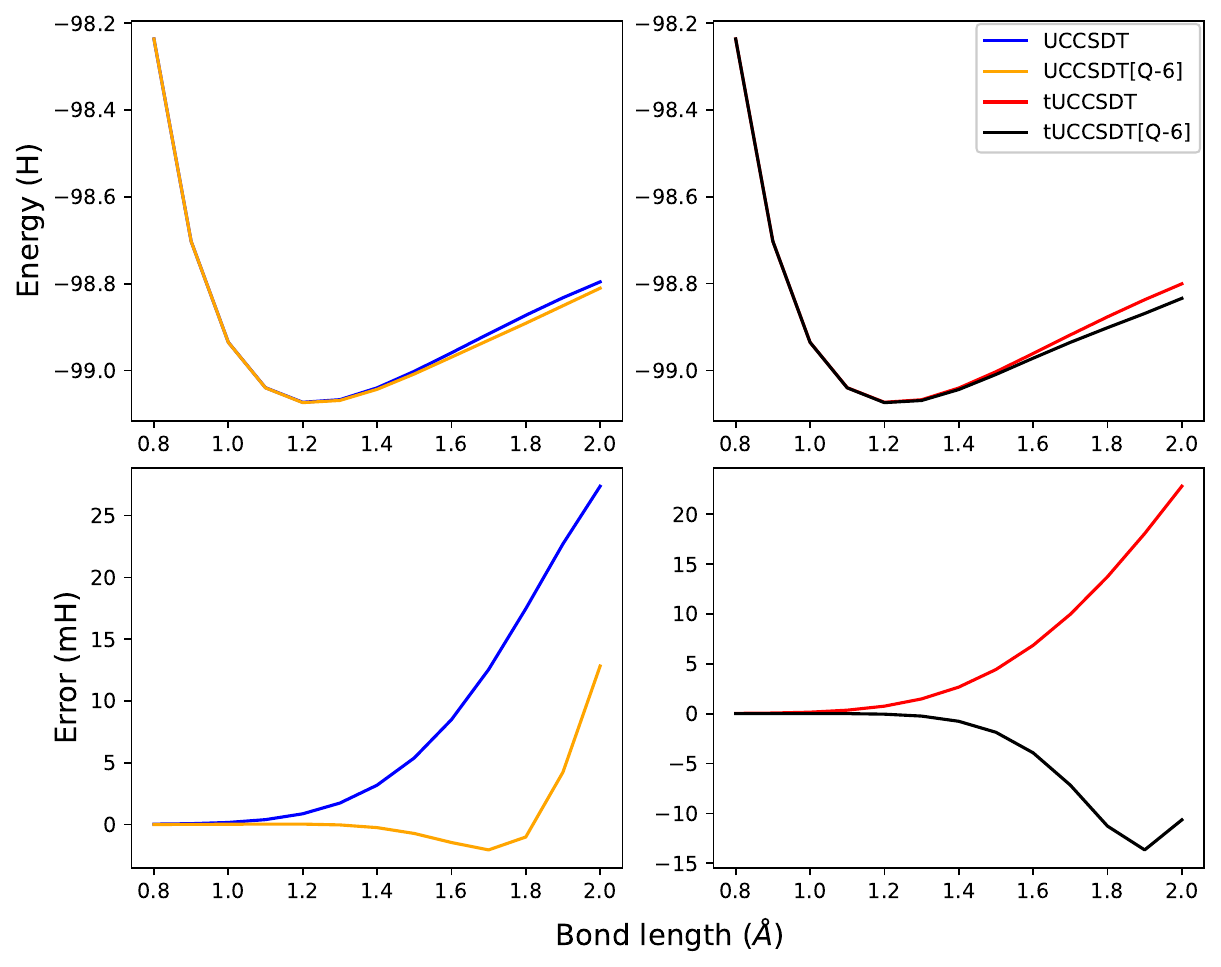}
 \caption{Potential energy surfaces computed with UCCSDT/[Q-6] and tUCCSDT/[Q-6] and corresponding errors with respect to FCI for the dissociation of BO$^-$ in the lowest energy, $^1\Sigma^+$ excited state.}
 \label{fig:BOminuspes}
\end{figure}

Turning to the $^1\Sigma^+$ excited state PES for BO$^-$, Figure \ref{fig:BOminuspes} again highlights the benefits of incorporating the [Q-6] correction. This particular system has been examined previously using single-reference techniques,\cite{peterson1989ground,zhai2007vibrationally} which have been shown to yield predictions that closely coincide with experiment.\cite{zhai2007vibrationally} Although this seems to suggest a single-reference determinant is sufficient at describing its electronic structure, we nevertheless found evidence of appreciable higher-order excitation effects for the $^1\Sigma^+$ excited state of BO$^-$ in a minimal basis set, as illustrated in Figure \ref{fig:BOminuspes}.

These trends are quantified in Table \ref{tab:errors}, which illustrate the [Q-6] correction reducing the MUE of t/UCCSDT from $>$ 7 mH to roughly 1 mH. This highlights the efficacy of the [Q-6] correction, and further shows that t/UCCSDT[Q-6] can achieve a chemically accurate MUE while simultaneously reducing the NPE of t/UCCSDT by roughly half. Analyzing these results in further detail, we found that both UCCSDT ansatze exhibit similar maximum errors of ~ 28 mH across the domain of this PES, which is quite large considering the assumed simplicity of the electronic structure. Nevertheless, the [Q-6] correction reduces this by half to 12.8-13.8 mH for the full and trotterized ansatze, respectively. 

\subsection{N$_2$}

\begin{figure}[ht!]
 \centering
 \includegraphics[width=\columnwidth]{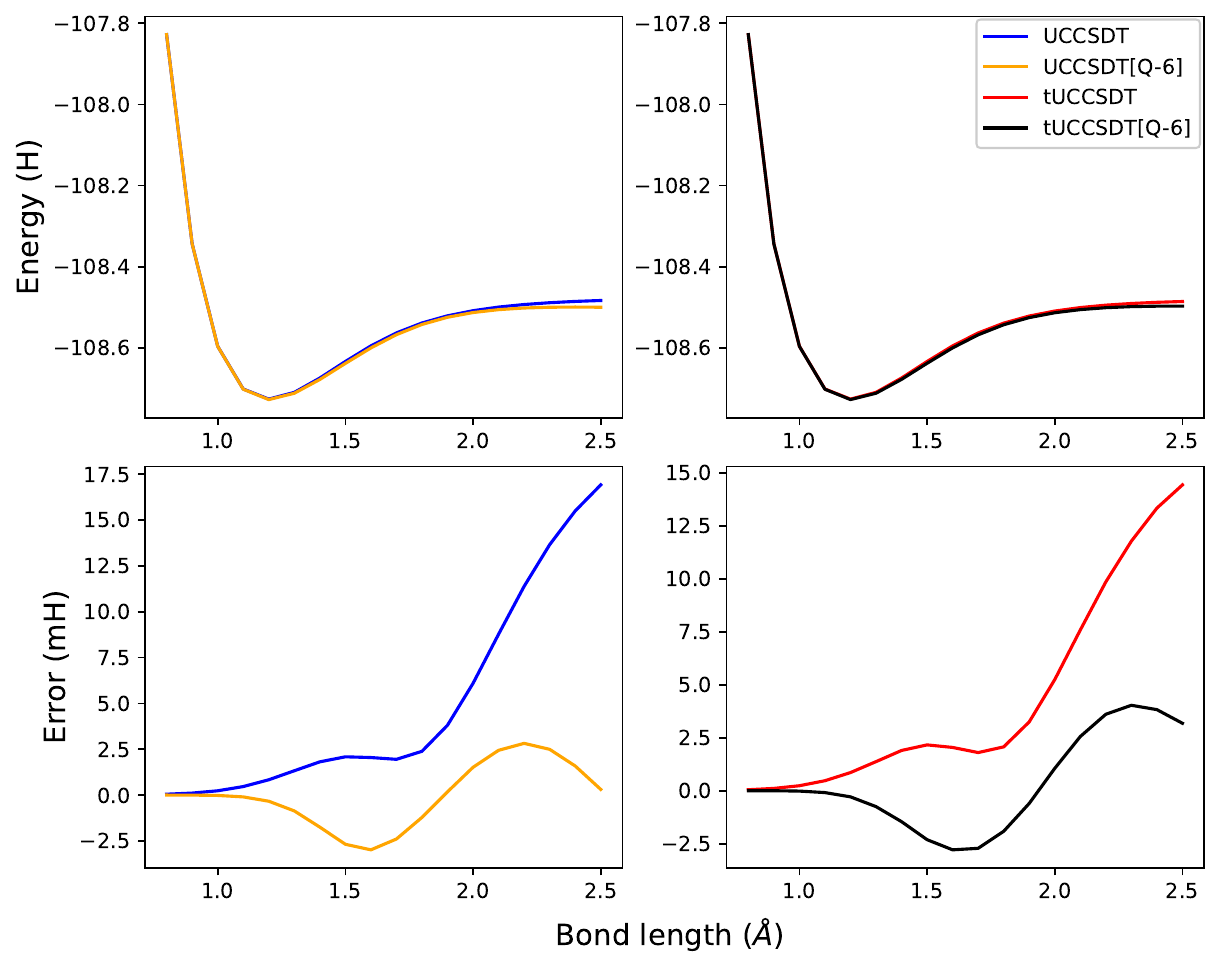}
 \caption{Potential energy surfaces computed with UCCSDT/[Q-6] and tUCCSDT/[Q-6] and corresponding errors with respect to FCI for the dissociation of N$_2$ in the lowest energy, singlet ground state.}
 \label{fig:N2pes}
\end{figure}

Breaking the triple bond N$_2$ represents a pathological problem for most electronic structure methods, especially those based on a single Slater determinant. Figure \ref{fig:N2pes} illustrates the t/UCCSDT results for this PES in addition to the [Q-6] correction. By adding explicit triple excitation operators to the ansatz, the overall error of t/UCCSDT with respect to FCI is improved as compared to the results in prior work involving the t/UCCSD ansatz\cite{windom2024new} which are further corroborated in Table \ref{tab:errors}. After 2.0\AA, however, even the addition of full triples excitation operators is not sufficient to manage the error, which exceed 13 mH at the worst point. In this region, the impact of implicit quadruples effects via [Q-6] are dramatic as shown in Figure \ref{fig:N2pes}. The maximum error of the [Q-6], based on the full and trotterized operator, are 2.8 and 4.0 mH, respectively. General trends are captured in Table \ref{tab:errors}, showing that t/UCCSDT[Q-6] reduces the MUE of t/UCCSDT from more than 4 mH to less than 1 mH and reduces the latter's NPE by more than half. 


\subsection{O$_2$}
\begin{figure}[ht!]
 \centering
 \includegraphics[width=\columnwidth]{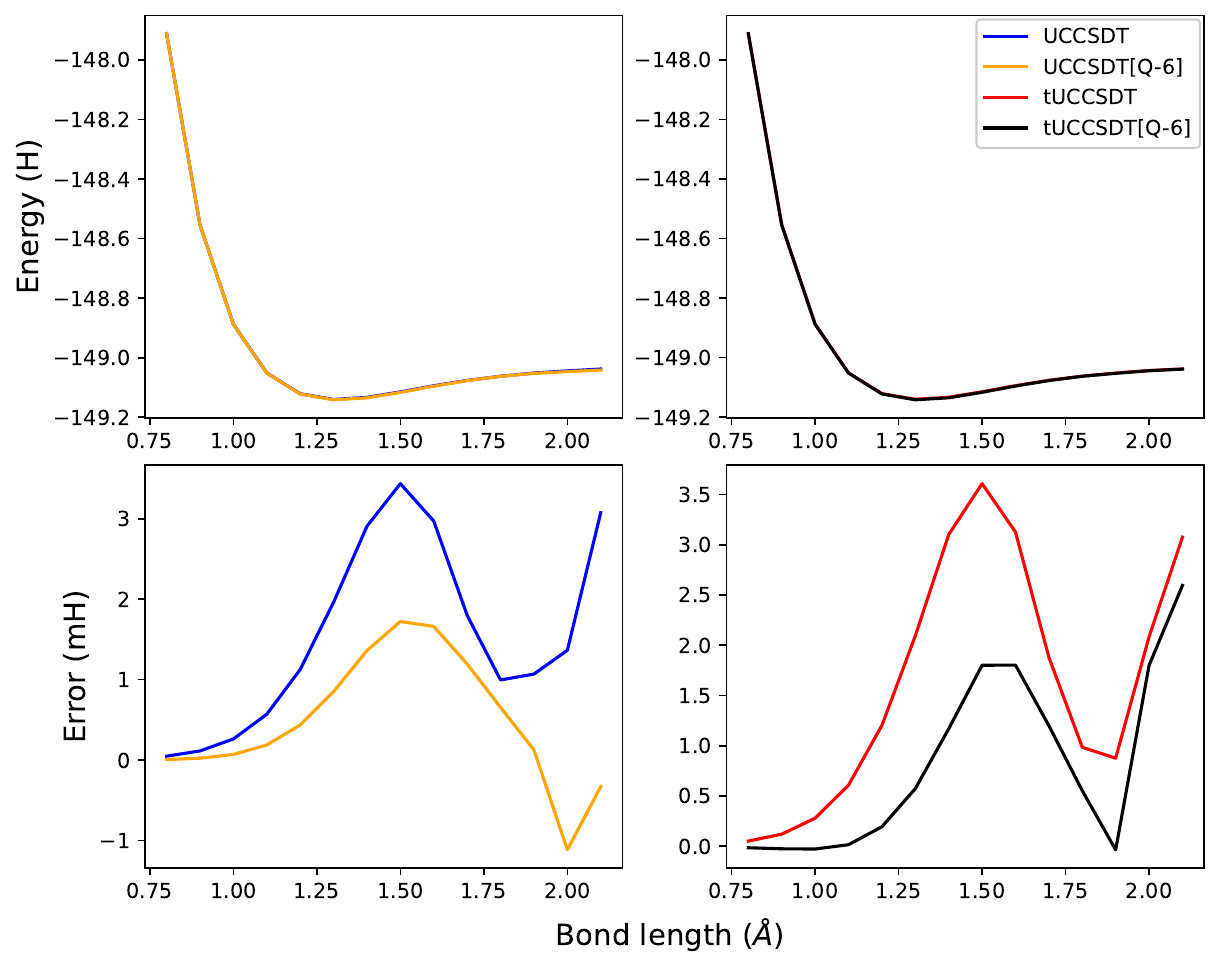}
 \caption{Potential energy surfaces computed with UCCSDT/[Q-6] and tUCCSDT/[Q-6] and corresponding errors with respect to FCI for the dissociation of $^1\Delta$ state of O$_2$.}
 \label{fig:O2pes}
\end{figure}

Similar to NF, O$_2$ has two low-lying singlet excited states: a$^1\Delta$ and b$^1\Sigma^+$. Prior work by the authors  found that minimal basis set UCCSD calculations on the a$^1\Delta$ electronic state in particular were in error with respect to FCI by about 9 mH.\cite{windom2024new} Figure \ref{fig:O2pes} shows that while improvements to prior t/UCCSD results can be made via explicit inclusion of infinite-order triples excitation operators, baseline t/UCCSDT alone neglects some electron correlation effects.  In terms of the maximum error, t/UCCSDT  significantly reduces the maximum t/UCCSD error reported in Ref.\citenum{windom2024new} to roughly 3-4 mH with respect to FCI. Nevertheless, adding the [Q-6] correction to both ansatz generally results in a more accurate description of the PES. To this point, the maximum error of UCCSDT[Q-6] with respect to FCI is reduced to 1.7 mH in the range between 0.8-1.8 \AA, while tUCCSDT[Q-6] yields a maximum error of 1.3 mH. The general trends illustrated in Figure \ref{fig:O2pes} result in t/UCCSDT[Q-6] achieving a chemically accurate MUE as recorded in Table \ref{tab:errors}. 

\subsection{Diagrammatic analysis of [Q-6]}

To pinpoint which terms in [Q-6] may be sensitive and/or dominate in different sections of the PES, we can decompose the overall [Q-6] correction into its constituent diagrammatic contributions, as illustrated in Figure \ref{fig:Qcorr}. In doing so, we can broadly associate particular diagrams with particular ``types" of net-quadruple excitation. For example, Diagram A depends purely on net-quadruple excitations out of the reference space using only triples cluster operators, whereas Diagram D depends purely on net-quadruple excitations out of the reference space - modulated by two, double electron excitations - via $T_2^2$. Diagrams B and C, then, consider correlation effects involving net-four electron excitations out of the reference space  quadruples that are modulated by coupling triply excited determinants (via $T_3$) with two, double electron excitations out of the reference (via $T_2^2$). 

\begin{figure}[ht!]
 \centering
 \includegraphics[width=\columnwidth]{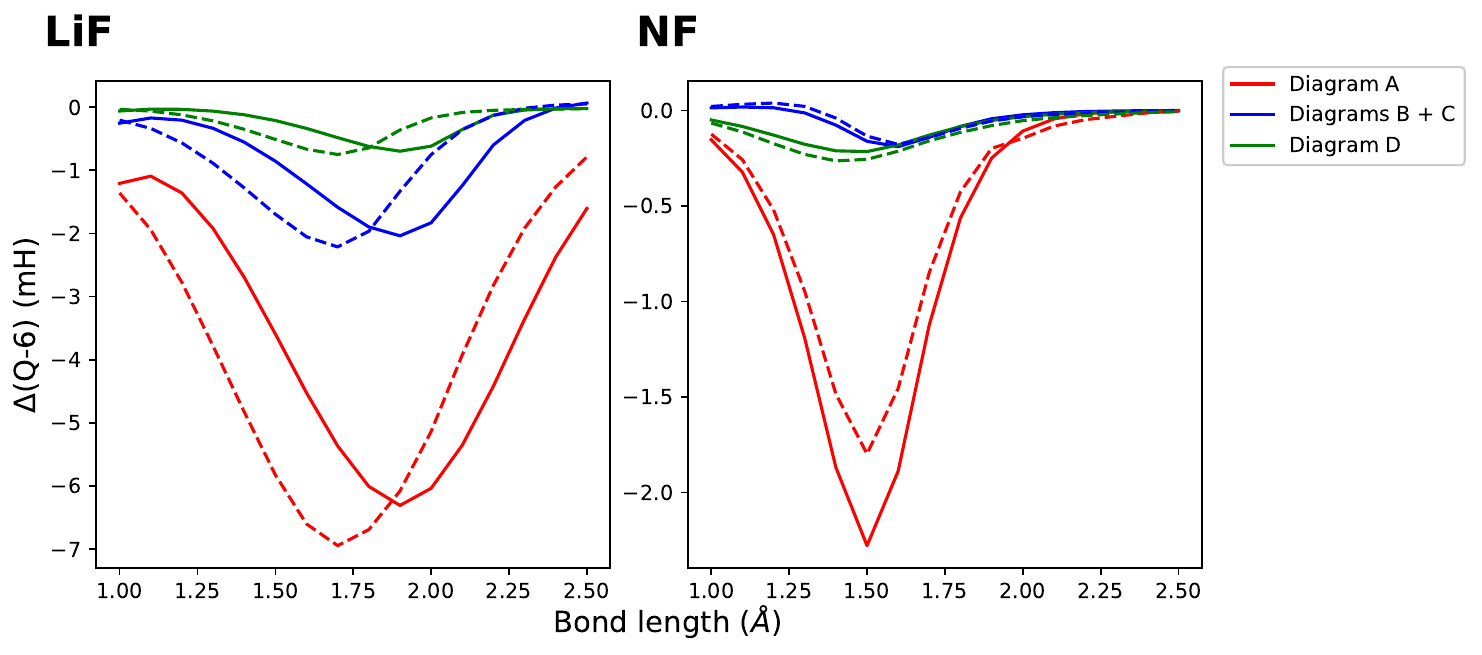}
 \caption{Decomposition of the [Q-6] correction according to the diagrams illustrated in Figure \ref{fig:Qcorr}. Solid lines denote individual diagrammatic contributions to the [Q-6] to UCCSDT, whereas dashed lines refer to tUCCSDT employing the ``default" operator ordering.}
 \label{fig:Q6contribs_wellbehaved}
\end{figure}

 Figure \ref{fig:Q6contribs_wellbehaved} depicts the individual diagrammatic contributions governing the [Q-6] correction for the LiF and NF examples with respect to both  the trotterized and full UCCSDT ansatz. We note that the dominant term in both examples, and for both flavors of UCC ansatz, is Diagram A which contributes the bulk of the [Q-6] energy correction to UCCSDT at around 6.5-7 and 1.7-2.5 mH for LiF and NF, respectively. In the case of LiF, Diagrams B and C are of secondary importance, yet still contribute roughly 2 mH toward the overall [Q-6] correction for both t/UCCSDT ansatz. However, these diagrams do not significantly participate in [Q-6] for NF. For both molecules, Diagram D marginally influences the net [Q-6] correction. Broadly speaking, these trends suggest that triply excited determinants are more important to the overall [Q-6] correction for systems that might conventionally be thought of as ``well-behaved" (e.g. dominated by dynamic correlation). The differences in [Q-6] using the trotterized and full UCCSDT operator are minimally shifted in some instances, but exhibit good agreement overall. In the following examples -  which could be conceived as being more pathological (e.g. influenced more by static and/or non-dynamic correlations) - we point out that the agreement between full and trotterized [Q-6] was used as a metric to determine the operator ordering defining the trotterized ansatz; additional information about this issue is further discussed in the Supplementary Material. 

\begin{figure}[ht!]
 \centering
 \includegraphics[width=\columnwidth]{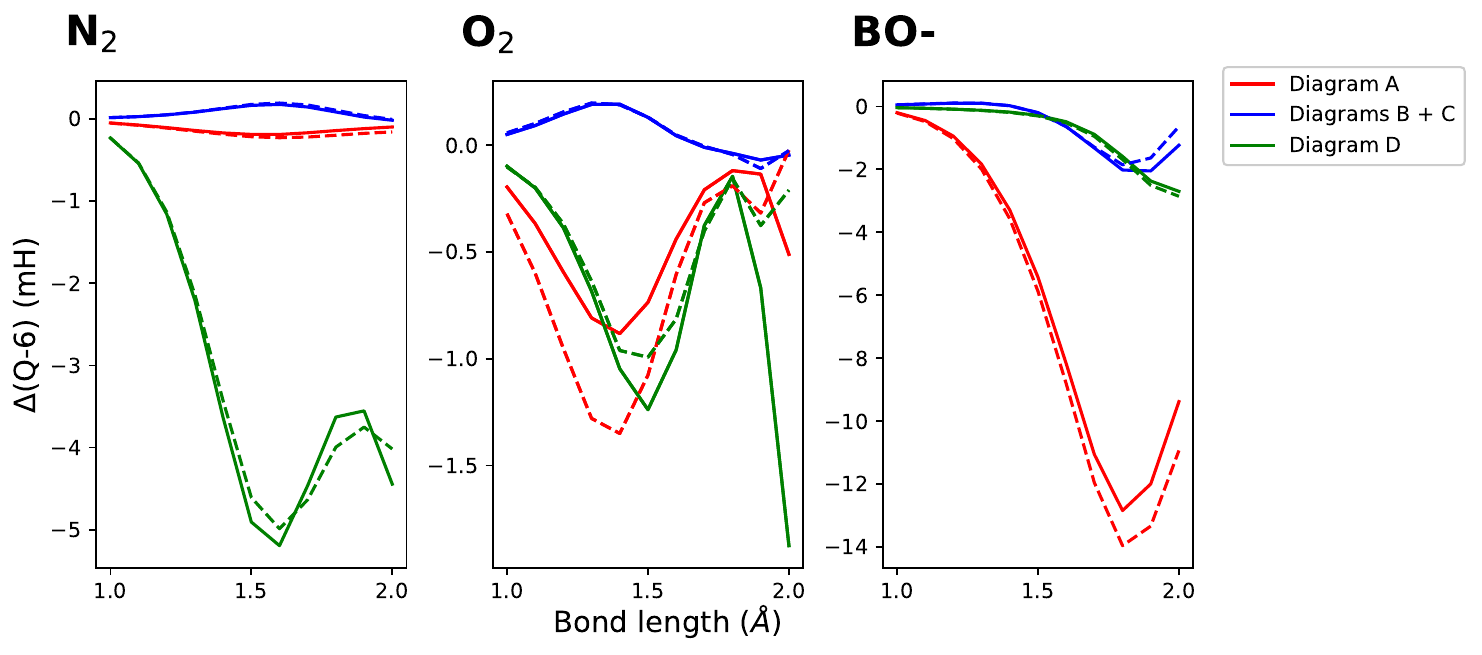}
 \caption{Decomposition of the [Q-6] correction according to the diagrams illustrated in Figure \ref{fig:Qcorr}. Solid lines denote individual diagrammatic contributions to the [Q-6] to  UCCSDT, whereas dashed lines refer to tUCCSDT employing the ``reverse" operator ordering.}
 \label{fig:Q6contribs_bad}
\end{figure}

Next, our analysis shifts to the N$_2$, O$_2$, and BO$^-$ PESs which -- conventionally speaking -- represent more pathological examples. As previously done, Figure \ref{fig:Q6contribs_bad} decomposes the [Q-6] correction into diagrammatic contributions for these molecules. For N$_2$ and O$_2$, Diagram D seems to dominate the overall [Q-6] which  is more pronounced in the case of N$_2$. This makes intuitive sense as these examples involve the breaking of $n>1$ chemical bonds at a time, which inherently involve the excitation of $2n$ electrons out of the reference space; these effects are naturally encapsulated by net-quadruple excitation effects involving $W_NT_2^2$. Unlike the correction for N$_2$ which depends almost entirely on diagram D, the correction for O$_2$ tends to rely - almost equally - on diagram A near the equilibrium. In contrast to these two examples, diagram A appears to dominate the correction for the BO$^-$ molecule, which is particularly notable in stretched regimes. This is somewhat counter-intuitive since BO$^-$ is isoelectronic with N$_2$, yet the ``important" diagrammatic contributions governing the overall [Q-6] correction in both species is fundamentally different. Of course, this is not necessarily a fair one-to-one comparison since these molecules would exhibit fundamentally different chemically bonding characteristics, with BO$^-$ being more amenable to ionic bonding due to the induced dipole moment whereas N$_2$ would form a purely covalent bond. Regardless,  diagrams B and C minimally contribute to the [Q-6] correction overall. Here again we note that [Q-6] corrections based on trotterization largely follow the corresponding full operator's behavior, which is a characteristic that should be obeyed by rigorously equivalent theories.


\section{Conclusion}
\label{sec:conclusion}

In this work, we propose a \textit{post hoc} perturbative correction to the UCCSDT ansatz that considers the leading-order effects associated with neglected quadruple excitations, and which is correct through sixth-order in MBPT  referred to as [Q-6]. We present two different ways to derive the [Q-6] correction, both of which use UCCSDT as the ``zeroth-order" reference wavefunction. The first takes advantage of the perturbation partitioning technique --originally proposed by Lowdin -- while the second considers the relationship between terms of the UCC energy functional which are correct through sixth-order in MBPT and the corresponding set of residual equations.  The similarities and differences of  perturbative quadruples corrections in conventional CC and UCC theories are also discussed. 

The efficacy of the [Q-6] correction to t/UCCSDT is assessed by examining the potential energy surfaces (PESs) of small diatomic molecules - LiF, NF, BO$^-$, N$_2$, and O$_2$ - which were chosen based on the importance of higher-order excitation effects in resolving the FCI within a minimal basis set. We found that while the t/UCCSDT ansatz is routinely capable of chemical accuracy within equilibrium regions, both ansatz exhibit errors - sometimes significantly - larger than 1 mH in regions outside equilibrium. By augmenting the t/UCCSDT method with the [Q-6] correction, errors consistently improve with respect to FCI across the majority of the PESs sampled in this work. In particular regions (e.g. equilibrium vs stretched) of the PES, the margin of improvement offered by t/UCCSDT[Q-6] can be quite significant, ranging from less than 1 mH to several orders of magnitude in error improvement over baseline t/UCCSDT.  We find that in all cases, t/UCCSDT[Q-6] achieves a MUE that is either chemically accurate, or exceedingly close to being chemically accurate, and significantly reduces the NPE as compared to t/UCCSDT. Future work will explore ways to construct perturbative corrections in UCC theory that retain the underlying ansatz's variational character, and are designed to further reduce the algorithmic complexity of the classically-computed, [Q-6] correction.

We further emphasize that, in the context of quantum computing, constructing the [Q-6] correction does not require additional quantum resources beyond what is required to perform the baseline t/UCCSDT calculation since it can be computed on a classical computer using an $\mathcal{O}(N^9)$  algorithm. We believe the current work provides additional evidence of the potential benefits in adopting a hybrid-compute workflow, which partitions the work to recover electron correlation in a way that intelligently leverages the strengths of existing classical and quantum computing devices in tandem, while minimizing their limitations.

\section{Acknowledgments}
 Z.W.W. and D.C. acknowledge support by the “Embedding Quantum Computing into Many-body Frameworks for Strongly Correlated Molecular and Materials Systems” project, which is funded by the U.S. Department of Energy (DOE), Office of Science, Office of Basic Energy Sciences, the Division of Chemical Sciences, Geosciences, and Biosciences. L.B. acknowledges support from the Laboratory Directed Research and Development Program of Oak Ridge National Laboratory, managed by UT-Battelle, LLC, for the US Department
of Energy. R.J.B acknowledges support from the Air Force Office of Scientific Research under AFOSR Award No. FA9550-23-1-0118. This research used resources of the Compute and Data Environment for Science (CADES) at the Oak Ridge National Laboratory, which is supported by the Office of Science of the U.S. Department of Energy under Contract No. DE-AC05-00OR22725.




\bibliography{biblio}

\providecommand{\noopsort}[1]{}\providecommand{\singleletter}[1]{#1}%
\begin{thebibliography}{3}
\expandafter\ifx\csname natexlab\endcsname\relax\def\natexlab#1{#1}\fi
\expandafter\ifx\csname bibnamefont\endcsname\relax
  \def\bibnamefont#1{#1}\fi
\expandafter\ifx\csname bibfnamefont\endcsname\relax
  \def\bibfnamefont#1{#1}\fi
\expandafter\ifx\csname citenamefont\endcsname\relax
  \def\citenamefont#1{#1}\fi
\expandafter\ifx\csname url\endcsname\relax
  \def\url#1{\texttt{#1}}\fi
\expandafter\ifx\csname urlprefix\endcsname\relax\def\urlprefix{URL }\fi
\providecommand{\bibinfo}[2]{#2}
\providecommand{\eprint}[2][]{\url{#2}}

\bibitem[{\citenamefont{Evangelista et~al.}(2019)\citenamefont{Evangelista, Chan, and Scuseria}}]{evangelista2019exact}
\bibinfo{author}{\bibfnamefont{F.~A.} \bibnamefont{Evangelista}}, \bibinfo{author}{\bibfnamefont{G.~K.} \bibnamefont{Chan}}, \bibnamefont{and} \bibinfo{author}{\bibfnamefont{G.~E.} \bibnamefont{Scuseria}}, \bibinfo{journal}{The Journal of Chemical Physics} \textbf{\bibinfo{volume}{151}}, \bibinfo{pages}{244112} (\bibinfo{year}{2019}).

\bibitem[{\citenamefont{Grimsley et~al.}(2019)\citenamefont{Grimsley, Claudino, Economou, Barnes, and Mayhall}}]{grimsley2019trotterized}
\bibinfo{author}{\bibfnamefont{H.~R.} \bibnamefont{Grimsley}}, \bibinfo{author}{\bibfnamefont{D.}~\bibnamefont{Claudino}}, \bibinfo{author}{\bibfnamefont{S.~E.} \bibnamefont{Economou}}, \bibinfo{author}{\bibfnamefont{E.}~\bibnamefont{Barnes}}, \bibnamefont{and} \bibinfo{author}{\bibfnamefont{N.~J.} \bibnamefont{Mayhall}}, \bibinfo{journal}{Journal of Chemical Theory and Computation} \textbf{\bibinfo{volume}{16}}, \bibinfo{pages}{1} (\bibinfo{year}{2019}).

\bibitem[{\citenamefont{Windom et~al.}(2024)\citenamefont{Windom, Claudino, and Bartlett}}]{windom2024new}
\bibinfo{author}{\bibfnamefont{Z.~W.} \bibnamefont{Windom}}, \bibinfo{author}{\bibfnamefont{D.}~\bibnamefont{Claudino}}, \bibnamefont{and} \bibinfo{author}{\bibfnamefont{R.~J.} \bibnamefont{Bartlett}}, \bibinfo{journal}{The Journal of Chemical Physics} \textbf{\bibinfo{volume}{160}} (\bibinfo{year}{2024}).

\end{thebibliography}


\providecommand{\noopsort}[1]{}\providecommand{\singleletter}[1]{#1}%
\begin{thebibliography}{65}
\expandafter\ifx\csname natexlab\endcsname\relax\def\natexlab#1{#1}\fi
\expandafter\ifx\csname bibnamefont\endcsname\relax
  \def\bibnamefont#1{#1}\fi
\expandafter\ifx\csname bibfnamefont\endcsname\relax
  \def\bibfnamefont#1{#1}\fi
\expandafter\ifx\csname citenamefont\endcsname\relax
  \def\citenamefont#1{#1}\fi
\expandafter\ifx\csname url\endcsname\relax
  \def\url#1{\texttt{#1}}\fi
\expandafter\ifx\csname urlprefix\endcsname\relax\def\urlprefix{URL }\fi
\providecommand{\bibinfo}[2]{#2}
\providecommand{\eprint}[2][]{\url{#2}}

\bibitem[{\citenamefont{Bartlett}(1989)}]{bartlett1989coupled}
\bibinfo{author}{\bibfnamefont{R.~J.} \bibnamefont{Bartlett}}, \bibinfo{journal}{The Journal of Physical Chemistry} \textbf{\bibinfo{volume}{93}}, \bibinfo{pages}{1697} (\bibinfo{year}{1989}).

\bibitem[{\citenamefont{Bartlett}(1981)}]{bartlett1981many}
\bibinfo{author}{\bibfnamefont{R.~J.} \bibnamefont{Bartlett}}, \bibinfo{journal}{Annual Review of Physical Chemistry} \textbf{\bibinfo{volume}{32}}, \bibinfo{pages}{359} (\bibinfo{year}{1981}).

\bibitem[{\citenamefont{Lengsfield et~al.}(1984)\citenamefont{Lengsfield, Saxe, and Yarkony}}]{lengsfield1984evaluation}
\bibinfo{author}{\bibfnamefont{B.~H.} \bibnamefont{Lengsfield}}, \bibinfo{author}{\bibfnamefont{P.}~\bibnamefont{Saxe}}, \bibnamefont{and} \bibinfo{author}{\bibfnamefont{D.~R.} \bibnamefont{Yarkony}}, \bibinfo{journal}{The Journal of Chemical Physics} \textbf{\bibinfo{volume}{81}}, \bibinfo{pages}{4549} (\bibinfo{year}{1984}).

\bibitem[{\citenamefont{Woywod et~al.}(1994)\citenamefont{Woywod, Domcke, Sobolewski, and Werner}}]{woywod1994characterization}
\bibinfo{author}{\bibfnamefont{C.}~\bibnamefont{Woywod}}, \bibinfo{author}{\bibfnamefont{W.}~\bibnamefont{Domcke}}, \bibinfo{author}{\bibfnamefont{A.~L.} \bibnamefont{Sobolewski}}, \bibnamefont{and} \bibinfo{author}{\bibfnamefont{H.-J.} \bibnamefont{Werner}}, \bibinfo{journal}{The Journal of Chemical Physics} \textbf{\bibinfo{volume}{100}}, \bibinfo{pages}{1400} (\bibinfo{year}{1994}).

\bibitem[{\citenamefont{Bartlett and Musia{\l}}(2007)}]{bartlett2007coupled}
\bibinfo{author}{\bibfnamefont{R.~J.} \bibnamefont{Bartlett}} \bibnamefont{and} \bibinfo{author}{\bibfnamefont{M.}~\bibnamefont{Musia{\l}}}, \bibinfo{journal}{Reviews of Modern Physics} \textbf{\bibinfo{volume}{79}}, \bibinfo{pages}{291} (\bibinfo{year}{2007}).

\bibitem[{\citenamefont{Shavitt and Bartlett}(2009)}]{shavitt2009many}
\bibinfo{author}{\bibfnamefont{I.}~\bibnamefont{Shavitt}} \bibnamefont{and} \bibinfo{author}{\bibfnamefont{R.~J.} \bibnamefont{Bartlett}}, \emph{\bibinfo{title}{{Many-body methods in chemistry and physics: MBPT and coupled-cluster theory}}} (\bibinfo{publisher}{Cambridge University Press}, \bibinfo{year}{2009}).

\bibitem[{\citenamefont{Paldus et~al.}(1984{\natexlab{a}})\citenamefont{Paldus, {\v{C}}{\'\i}{\v{z}}ek, and Takahashi}}]{paldus1984approximate}
\bibinfo{author}{\bibfnamefont{J.}~\bibnamefont{Paldus}}, \bibinfo{author}{\bibfnamefont{J.}~\bibnamefont{{\v{C}}{\'\i}{\v{z}}ek}}, \bibnamefont{and} \bibinfo{author}{\bibfnamefont{M.}~\bibnamefont{Takahashi}}, \bibinfo{journal}{Physical Review A} \textbf{\bibinfo{volume}{30}}, \bibinfo{pages}{2193} (\bibinfo{year}{1984}{\natexlab{a}}).

\bibitem[{\citenamefont{Chan et~al.}(2004)\citenamefont{Chan, K{\'a}llay, and Gauss}}]{chan2004state}
\bibinfo{author}{\bibfnamefont{G.~K.-L.} \bibnamefont{Chan}}, \bibinfo{author}{\bibfnamefont{M.}~\bibnamefont{K{\'a}llay}}, \bibnamefont{and} \bibinfo{author}{\bibfnamefont{J.}~\bibnamefont{Gauss}}, \bibinfo{journal}{The Journal of Chemical Physics} \textbf{\bibinfo{volume}{121}}, \bibinfo{pages}{6110} (\bibinfo{year}{2004}).

\bibitem[{\citenamefont{Bartlett and Purvis}(1978)}]{bartlett1978many}
\bibinfo{author}{\bibfnamefont{R.~J.} \bibnamefont{Bartlett}} \bibnamefont{and} \bibinfo{author}{\bibfnamefont{G.~D.} \bibnamefont{Purvis}}, \bibinfo{journal}{International Journal of Quantum Chemistry} \textbf{\bibinfo{volume}{14}}, \bibinfo{pages}{561} (\bibinfo{year}{1978}).

\bibitem[{\citenamefont{Bartlett et~al.}(1979)\citenamefont{Bartlett, Shavitt, and Purvis~III}}]{bartlett1979quartic}
\bibinfo{author}{\bibfnamefont{R.~J.} \bibnamefont{Bartlett}}, \bibinfo{author}{\bibfnamefont{I.}~\bibnamefont{Shavitt}}, \bibnamefont{and} \bibinfo{author}{\bibfnamefont{G.~D.} \bibnamefont{Purvis~III}}, \bibinfo{journal}{The Journal of Chemical Physics} \textbf{\bibinfo{volume}{71}}, \bibinfo{pages}{281} (\bibinfo{year}{1979}).

\bibitem[{\citenamefont{Kucharski and Bartlett}(1989)}]{kucharski1989coupled}
\bibinfo{author}{\bibfnamefont{S.~A.} \bibnamefont{Kucharski}} \bibnamefont{and} \bibinfo{author}{\bibfnamefont{R.~J.} \bibnamefont{Bartlett}}, \bibinfo{journal}{Chemical Physics Letters} \textbf{\bibinfo{volume}{158}}, \bibinfo{pages}{550} (\bibinfo{year}{1989}).

\bibitem[{\citenamefont{Windom et~al.}(2024{\natexlab{a}})\citenamefont{Windom, Perera, and Bartlett}}]{windom2024ultimate}
\bibinfo{author}{\bibfnamefont{Z.~W.} \bibnamefont{Windom}}, \bibinfo{author}{\bibfnamefont{A.}~\bibnamefont{Perera}}, \bibnamefont{and} \bibinfo{author}{\bibfnamefont{R.~J.} \bibnamefont{Bartlett}}, \bibinfo{journal}{The Journal of Chemical Physics} \textbf{\bibinfo{volume}{161}} (\bibinfo{year}{2024}{\natexlab{a}}).

\bibitem[{\citenamefont{Windom and Bartlett}(2024)}]{windoma2024t2}
\bibinfo{author}{\bibfnamefont{Z.~W.} \bibnamefont{Windom}} \bibnamefont{and} \bibinfo{author}{\bibfnamefont{R.~J.} \bibnamefont{Bartlett}}, \emph{\bibinfo{title}{An effective, “strong” electron correlation estimate using alternative formulations of T2-based coupled cluster theory and the factorization theorem}} (\bibinfo{publisher}{Elsevier}, \bibinfo{year}{2024}), p. \bibinfo{pages}{1–14}.

\bibitem[{\citenamefont{Paldus and Boyle}(1982)}]{paldus1982cluster}
\bibinfo{author}{\bibfnamefont{J.}~\bibnamefont{Paldus}} \bibnamefont{and} \bibinfo{author}{\bibfnamefont{M.}~\bibnamefont{Boyle}}, \bibinfo{journal}{International Journal of Quantum Chemistry} \textbf{\bibinfo{volume}{22}}, \bibinfo{pages}{1281} (\bibinfo{year}{1982}).

\bibitem[{\citenamefont{Paldus et~al.}(1984{\natexlab{b}})\citenamefont{Paldus, Takahashi, and Cho}}]{paldus1984degeneracy}
\bibinfo{author}{\bibfnamefont{J.}~\bibnamefont{Paldus}}, \bibinfo{author}{\bibfnamefont{M.}~\bibnamefont{Takahashi}}, \bibnamefont{and} \bibinfo{author}{\bibfnamefont{B.}~\bibnamefont{Cho}}, \bibinfo{journal}{International Journal of Quantum Chemistry} \textbf{\bibinfo{volume}{26}}, \bibinfo{pages}{237} (\bibinfo{year}{1984}{\natexlab{b}}).

\bibitem[{\citenamefont{Piecuch et~al.}(1990)\citenamefont{Piecuch, Zarrabian, Paldus, and {\v{C}}{\'\i}{\v{z}}ek}}]{piecuch1990coupled}
\bibinfo{author}{\bibfnamefont{P.}~\bibnamefont{Piecuch}}, \bibinfo{author}{\bibfnamefont{S.}~\bibnamefont{Zarrabian}}, \bibinfo{author}{\bibfnamefont{J.}~\bibnamefont{Paldus}}, \bibnamefont{and} \bibinfo{author}{\bibfnamefont{J.}~\bibnamefont{{\v{C}}{\'\i}{\v{z}}ek}}, \bibinfo{journal}{Physical Review B} \textbf{\bibinfo{volume}{42}}, \bibinfo{pages}{3351} (\bibinfo{year}{1990}).

\bibitem[{\citenamefont{Tajti et~al.}(2004)\citenamefont{Tajti, Szalay, Cs{\'a}sz{\'a}r, K{\'a}llay, Gauss, Valeev, Flowers, V{\'a}zquez, and Stanton}}]{tajti2004heat}
\bibinfo{author}{\bibfnamefont{A.}~\bibnamefont{Tajti}}, \bibinfo{author}{\bibfnamefont{P.~G.} \bibnamefont{Szalay}}, \bibinfo{author}{\bibfnamefont{A.~G.} \bibnamefont{Cs{\'a}sz{\'a}r}}, \bibinfo{author}{\bibfnamefont{M.}~\bibnamefont{K{\'a}llay}}, \bibinfo{author}{\bibfnamefont{J.}~\bibnamefont{Gauss}}, \bibinfo{author}{\bibfnamefont{E.~F.} \bibnamefont{Valeev}}, \bibinfo{author}{\bibfnamefont{B.~A.} \bibnamefont{Flowers}}, \bibinfo{author}{\bibfnamefont{J.}~\bibnamefont{V{\'a}zquez}}, \bibnamefont{and} \bibinfo{author}{\bibfnamefont{J.~F.} \bibnamefont{Stanton}}, \bibinfo{journal}{The Journal of Chemical Physics} \textbf{\bibinfo{volume}{121}}, \bibinfo{pages}{11599} (\bibinfo{year}{2004}).

\bibitem[{\citenamefont{Bomble et~al.}(2006)\citenamefont{Bomble, V{\'a}zquez, K{\'a}llay, Michauk, Szalay, Cs{\'a}sz{\'a}r, Gauss, and Stanton}}]{bomble2006high}
\bibinfo{author}{\bibfnamefont{Y.~J.} \bibnamefont{Bomble}}, \bibinfo{author}{\bibfnamefont{J.}~\bibnamefont{V{\'a}zquez}}, \bibinfo{author}{\bibfnamefont{M.}~\bibnamefont{K{\'a}llay}}, \bibinfo{author}{\bibfnamefont{C.}~\bibnamefont{Michauk}}, \bibinfo{author}{\bibfnamefont{P.~G.} \bibnamefont{Szalay}}, \bibinfo{author}{\bibfnamefont{A.~G.} \bibnamefont{Cs{\'a}sz{\'a}r}}, \bibinfo{author}{\bibfnamefont{J.}~\bibnamefont{Gauss}}, \bibnamefont{and} \bibinfo{author}{\bibfnamefont{J.~F.} \bibnamefont{Stanton}}, \bibinfo{journal}{The Journal of Chemical Physics} \textbf{\bibinfo{volume}{125}} (\bibinfo{year}{2006}).

\bibitem[{\citenamefont{Martin and de~Oliveira}(1999)}]{martin1999towards}
\bibinfo{author}{\bibfnamefont{J.~M.} \bibnamefont{Martin}} \bibnamefont{and} \bibinfo{author}{\bibfnamefont{G.}~\bibnamefont{de~Oliveira}}, \bibinfo{journal}{The Journal of Chemical Physics} \textbf{\bibinfo{volume}{111}}, \bibinfo{pages}{1843} (\bibinfo{year}{1999}).

\bibitem[{\citenamefont{Boese et~al.}(2004)\citenamefont{Boese, Oren, Atasoylu, Martin, K{\'a}llay, and Gauss}}]{boese2004w3}
\bibinfo{author}{\bibfnamefont{A.~D.} \bibnamefont{Boese}}, \bibinfo{author}{\bibfnamefont{M.}~\bibnamefont{Oren}}, \bibinfo{author}{\bibfnamefont{O.}~\bibnamefont{Atasoylu}}, \bibinfo{author}{\bibfnamefont{J.~M.} \bibnamefont{Martin}}, \bibinfo{author}{\bibfnamefont{M.}~\bibnamefont{K{\'a}llay}}, \bibnamefont{and} \bibinfo{author}{\bibfnamefont{J.}~\bibnamefont{Gauss}}, \bibinfo{journal}{The Journal of Chemical Physics} \textbf{\bibinfo{volume}{120}}, \bibinfo{pages}{4129} (\bibinfo{year}{2004}).

\bibitem[{\citenamefont{Karton et~al.}(2006)\citenamefont{Karton, Rabinovich, Martin, and Ruscic}}]{karton2006w4}
\bibinfo{author}{\bibfnamefont{A.}~\bibnamefont{Karton}}, \bibinfo{author}{\bibfnamefont{E.}~\bibnamefont{Rabinovich}}, \bibinfo{author}{\bibfnamefont{J.~M.} \bibnamefont{Martin}}, \bibnamefont{and} \bibinfo{author}{\bibfnamefont{B.}~\bibnamefont{Ruscic}}, \bibinfo{journal}{The Journal of Chemical Physics} \textbf{\bibinfo{volume}{125}} (\bibinfo{year}{2006}).

\bibitem[{\citenamefont{Kucharski and Bartlett}(1993)}]{kucharski1993coupled}
\bibinfo{author}{\bibfnamefont{S.~A.} \bibnamefont{Kucharski}} \bibnamefont{and} \bibinfo{author}{\bibfnamefont{R.~J.} \bibnamefont{Bartlett}}, \bibinfo{journal}{Chemical Physics Letters} \textbf{\bibinfo{volume}{206}}, \bibinfo{pages}{574} (\bibinfo{year}{1993}).

\bibitem[{\citenamefont{Bomble et~al.}(2005)\citenamefont{Bomble, Stanton, K{\'a}llay, and Gauss}}]{bomble2005coupled}
\bibinfo{author}{\bibfnamefont{Y.~J.} \bibnamefont{Bomble}}, \bibinfo{author}{\bibfnamefont{J.~F.} \bibnamefont{Stanton}}, \bibinfo{author}{\bibfnamefont{M.}~\bibnamefont{K{\'a}llay}}, \bibnamefont{and} \bibinfo{author}{\bibfnamefont{J.}~\bibnamefont{Gauss}}, \bibinfo{journal}{The Journal of Chemical Physics} \textbf{\bibinfo{volume}{123}} (\bibinfo{year}{2005}).

\bibitem[{\citenamefont{Kucharski and Bartlett}(1998{\natexlab{a}})}]{kucharski1998sixth}
\bibinfo{author}{\bibfnamefont{S.~A.} \bibnamefont{Kucharski}} \bibnamefont{and} \bibinfo{author}{\bibfnamefont{R.~J.} \bibnamefont{Bartlett}}, \bibinfo{journal}{The Journal of Chemical Physics} \textbf{\bibinfo{volume}{108}}, \bibinfo{pages}{5255} (\bibinfo{year}{1998}{\natexlab{a}}).

\bibitem[{\citenamefont{Kucharski and Bartlett}(1986)}]{kucharski1986fifth}
\bibinfo{author}{\bibfnamefont{S.~A.} \bibnamefont{Kucharski}} \bibnamefont{and} \bibinfo{author}{\bibfnamefont{R.~J.} \bibnamefont{Bartlett}}, in \emph{\bibinfo{booktitle}{Advances in Quantum Chemistry}} (\bibinfo{publisher}{Elsevier}, \bibinfo{year}{1986}), vol.~\bibinfo{volume}{18}, pp. \bibinfo{pages}{281--344}.

\bibitem[{\citenamefont{Kucharski and Bartlett}(1998{\natexlab{b}})}]{kucharski1998efficient}
\bibinfo{author}{\bibfnamefont{S.~A.} \bibnamefont{Kucharski}} \bibnamefont{and} \bibinfo{author}{\bibfnamefont{R.~J.} \bibnamefont{Bartlett}}, \bibinfo{journal}{The Journal of Chemical Physics} \textbf{\bibinfo{volume}{108}}, \bibinfo{pages}{9221} (\bibinfo{year}{1998}{\natexlab{b}}).

\bibitem[{\citenamefont{Thorpe et~al.}(2024)\citenamefont{Thorpe, Windom, Bartlett, and Matthews}}]{thorpe2024factorized}
\bibinfo{author}{\bibfnamefont{J.~H.} \bibnamefont{Thorpe}}, \bibinfo{author}{\bibfnamefont{Z.~W.} \bibnamefont{Windom}}, \bibinfo{author}{\bibfnamefont{R.~J.} \bibnamefont{Bartlett}}, \bibnamefont{and} \bibinfo{author}{\bibfnamefont{D.~A.} \bibnamefont{Matthews}}, \bibinfo{journal}{The Journal of Physical Chemistry A} \textbf{\bibinfo{volume}{128}}, \bibinfo{pages}{7720} (\bibinfo{year}{2024}).

\bibitem[{\citenamefont{Bartlett et~al.}(1989)\citenamefont{Bartlett, Kucharski, and Noga}}]{bartlett1989alternative}
\bibinfo{author}{\bibfnamefont{R.~J.} \bibnamefont{Bartlett}}, \bibinfo{author}{\bibfnamefont{S.~A.} \bibnamefont{Kucharski}}, \bibnamefont{and} \bibinfo{author}{\bibfnamefont{J.}~\bibnamefont{Noga}}, \bibinfo{journal}{Chemical Physics Letters} \textbf{\bibinfo{volume}{155}}, \bibinfo{pages}{133} (\bibinfo{year}{1989}).

\bibitem[{\citenamefont{Szalay et~al.}(1995)\citenamefont{Szalay, Nooijen, and Bartlett}}]{szalay1995alternative}
\bibinfo{author}{\bibfnamefont{P.~G.} \bibnamefont{Szalay}}, \bibinfo{author}{\bibfnamefont{M.}~\bibnamefont{Nooijen}}, \bibnamefont{and} \bibinfo{author}{\bibfnamefont{R.~J.} \bibnamefont{Bartlett}}, \bibinfo{journal}{The Journal of Chemical Physics} \textbf{\bibinfo{volume}{103}}, \bibinfo{pages}{281} (\bibinfo{year}{1995}).

\bibitem[{\citenamefont{Anand et~al.}(2022)\citenamefont{Anand, Schleich, Alperin-Lea, Jensen, Sim, D{\'\i}az-Tinoco, Kottmann, Degroote, Izmaylov, and Aspuru-Guzik}}]{anand2022quantum}
\bibinfo{author}{\bibfnamefont{A.}~\bibnamefont{Anand}}, \bibinfo{author}{\bibfnamefont{P.}~\bibnamefont{Schleich}}, \bibinfo{author}{\bibfnamefont{S.}~\bibnamefont{Alperin-Lea}}, \bibinfo{author}{\bibfnamefont{P.~W.} \bibnamefont{Jensen}}, \bibinfo{author}{\bibfnamefont{S.}~\bibnamefont{Sim}}, \bibinfo{author}{\bibfnamefont{M.}~\bibnamefont{D{\'\i}az-Tinoco}}, \bibinfo{author}{\bibfnamefont{J.~S.} \bibnamefont{Kottmann}}, \bibinfo{author}{\bibfnamefont{M.}~\bibnamefont{Degroote}}, \bibinfo{author}{\bibfnamefont{A.~F.} \bibnamefont{Izmaylov}}, \bibnamefont{and} \bibinfo{author}{\bibfnamefont{A.}~\bibnamefont{Aspuru-Guzik}}, \bibinfo{journal}{Chemical Society Reviews} \textbf{\bibinfo{volume}{51}}, \bibinfo{pages}{1659} (\bibinfo{year}{2022}).

\bibitem[{\citenamefont{Kutzelnigg and Koch}(1983)}]{kutzelnigg1983quantum}
\bibinfo{author}{\bibfnamefont{W.}~\bibnamefont{Kutzelnigg}} \bibnamefont{and} \bibinfo{author}{\bibfnamefont{S.}~\bibnamefont{Koch}}, \bibinfo{journal}{The Journal of Chemical Physics} \textbf{\bibinfo{volume}{79}}, \bibinfo{pages}{4315} (\bibinfo{year}{1983}).

\bibitem[{\citenamefont{Liu et~al.}(2018)\citenamefont{Liu, Asthana, Cheng, and Mukherjee}}]{liu2018unitary}
\bibinfo{author}{\bibfnamefont{J.}~\bibnamefont{Liu}}, \bibinfo{author}{\bibfnamefont{A.}~\bibnamefont{Asthana}}, \bibinfo{author}{\bibfnamefont{L.}~\bibnamefont{Cheng}}, \bibnamefont{and} \bibinfo{author}{\bibfnamefont{D.}~\bibnamefont{Mukherjee}}, \bibinfo{journal}{The Journal of Chemical Physics} \textbf{\bibinfo{volume}{148}} (\bibinfo{year}{2018}).

\bibitem[{\citenamefont{Hodecker et~al.}(2020)\citenamefont{Hodecker, Thielen, Liu, Rehn, and Dreuw}}]{hodecker2020third}
\bibinfo{author}{\bibfnamefont{M.}~\bibnamefont{Hodecker}}, \bibinfo{author}{\bibfnamefont{S.~M.} \bibnamefont{Thielen}}, \bibinfo{author}{\bibfnamefont{J.}~\bibnamefont{Liu}}, \bibinfo{author}{\bibfnamefont{D.~R.} \bibnamefont{Rehn}}, \bibnamefont{and} \bibinfo{author}{\bibfnamefont{A.}~\bibnamefont{Dreuw}}, \bibinfo{journal}{Journal of Chemical Theory and Computation} \textbf{\bibinfo{volume}{16}}, \bibinfo{pages}{3654} (\bibinfo{year}{2020}).

\bibitem[{\citenamefont{Liu and Cheng}(2021)}]{liu2021unitary}
\bibinfo{author}{\bibfnamefont{J.}~\bibnamefont{Liu}} \bibnamefont{and} \bibinfo{author}{\bibfnamefont{L.}~\bibnamefont{Cheng}}, \bibinfo{journal}{The Journal of Chemical Physics} \textbf{\bibinfo{volume}{155}} (\bibinfo{year}{2021}).

\bibitem[{\citenamefont{Liu et~al.}(2022)\citenamefont{Liu, Matthews, and Cheng}}]{liu2022quadratic}
\bibinfo{author}{\bibfnamefont{J.}~\bibnamefont{Liu}}, \bibinfo{author}{\bibfnamefont{D.~A.} \bibnamefont{Matthews}}, \bibnamefont{and} \bibinfo{author}{\bibfnamefont{L.}~\bibnamefont{Cheng}}, \bibinfo{journal}{Journal of Chemical Theory and Computation} \textbf{\bibinfo{volume}{18}}, \bibinfo{pages}{2281} (\bibinfo{year}{2022}).

\bibitem[{\citenamefont{Li et~al.}(2025)\citenamefont{Li, Zhang, Sheng, Gong, Chen, and Shuai}}]{li2025quantum}
\bibinfo{author}{\bibfnamefont{W.}~\bibnamefont{Li}}, \bibinfo{author}{\bibfnamefont{S.-X.} \bibnamefont{Zhang}}, \bibinfo{author}{\bibfnamefont{Z.}~\bibnamefont{Sheng}}, \bibinfo{author}{\bibfnamefont{C.}~\bibnamefont{Gong}}, \bibinfo{author}{\bibfnamefont{J.}~\bibnamefont{Chen}}, \bibnamefont{and} \bibinfo{author}{\bibfnamefont{Z.}~\bibnamefont{Shuai}}, \bibinfo{journal}{arXiv preprint arXiv:2501.04264}  (\bibinfo{year}{2025}).

\bibitem[{\citenamefont{Grimsley et~al.}(2019{\natexlab{a}})\citenamefont{Grimsley, Economou, Barnes, and Mayhall}}]{grimsley2019adaptive}
\bibinfo{author}{\bibfnamefont{H.~R.} \bibnamefont{Grimsley}}, \bibinfo{author}{\bibfnamefont{S.~E.} \bibnamefont{Economou}}, \bibinfo{author}{\bibfnamefont{E.}~\bibnamefont{Barnes}}, \bibnamefont{and} \bibinfo{author}{\bibfnamefont{N.~J.} \bibnamefont{Mayhall}}, \bibinfo{journal}{Nature Communications} \textbf{\bibinfo{volume}{10}}, \bibinfo{pages}{3007} (\bibinfo{year}{2019}{\natexlab{a}}).

\bibitem[{\citenamefont{Halder et~al.}(2024)\citenamefont{Halder, Mondal, and Maitra}}]{halder2024noise}
\bibinfo{author}{\bibfnamefont{D.}~\bibnamefont{Halder}}, \bibinfo{author}{\bibfnamefont{D.}~\bibnamefont{Mondal}}, \bibnamefont{and} \bibinfo{author}{\bibfnamefont{R.}~\bibnamefont{Maitra}}, \bibinfo{journal}{The Journal of Chemical Physics} \textbf{\bibinfo{volume}{160}} (\bibinfo{year}{2024}).

\bibitem[{\citenamefont{Windom et~al.}(2024{\natexlab{b}})\citenamefont{Windom, Claudino, and Bartlett}}]{windom2024attractive}
\bibinfo{author}{\bibfnamefont{Z.~W.} \bibnamefont{Windom}}, \bibinfo{author}{\bibfnamefont{D.}~\bibnamefont{Claudino}}, \bibnamefont{and} \bibinfo{author}{\bibfnamefont{R.~J.} \bibnamefont{Bartlett}}, \bibinfo{journal}{The Journal of Physical Chemistry A} \textbf{\bibinfo{volume}{128}}, \bibinfo{pages}{7036} (\bibinfo{year}{2024}{\natexlab{b}}).

\bibitem[{\citenamefont{Windom et~al.}(2024{\natexlab{c}})\citenamefont{Windom, Claudino, and Bartlett}}]{windom2024new}
\bibinfo{author}{\bibfnamefont{Z.~W.} \bibnamefont{Windom}}, \bibinfo{author}{\bibfnamefont{D.}~\bibnamefont{Claudino}}, \bibnamefont{and} \bibinfo{author}{\bibfnamefont{R.~J.} \bibnamefont{Bartlett}}, \bibinfo{journal}{The Journal of Chemical Physics} \textbf{\bibinfo{volume}{160}} (\bibinfo{year}{2024}{\natexlab{c}}).

\bibitem[{\citenamefont{Urban et~al.}(1985)\citenamefont{Urban, Noga, Cole, and Bartlett}}]{urban1985towards}
\bibinfo{author}{\bibfnamefont{M.}~\bibnamefont{Urban}}, \bibinfo{author}{\bibfnamefont{J.}~\bibnamefont{Noga}}, \bibinfo{author}{\bibfnamefont{S.~J.} \bibnamefont{Cole}}, \bibnamefont{and} \bibinfo{author}{\bibfnamefont{R.~J.} \bibnamefont{Bartlett}}, \bibinfo{journal}{The Journal of Chemical Physics} \textbf{\bibinfo{volume}{83}}, \bibinfo{pages}{4041} (\bibinfo{year}{1985}).

\bibitem[{\citenamefont{Raghavachari et~al.}(1989)\citenamefont{Raghavachari, Trucks, Pople, and Head-Gordon}}]{raghavachari1989fifth}
\bibinfo{author}{\bibfnamefont{K.}~\bibnamefont{Raghavachari}}, \bibinfo{author}{\bibfnamefont{G.~W.} \bibnamefont{Trucks}}, \bibinfo{author}{\bibfnamefont{J.~A.} \bibnamefont{Pople}}, \bibnamefont{and} \bibinfo{author}{\bibfnamefont{M.}~\bibnamefont{Head-Gordon}}, \bibinfo{journal}{Chemical Physics Letters} \textbf{\bibinfo{volume}{157}}, \bibinfo{pages}{479} (\bibinfo{year}{1989}).

\bibitem[{\citenamefont{Watts et~al.}(1993)\citenamefont{Watts, Gauss, and Bartlett}}]{watts1993coupled}
\bibinfo{author}{\bibfnamefont{J.~D.} \bibnamefont{Watts}}, \bibinfo{author}{\bibfnamefont{J.}~\bibnamefont{Gauss}}, \bibnamefont{and} \bibinfo{author}{\bibfnamefont{R.~J.} \bibnamefont{Bartlett}}, \bibinfo{journal}{The Journal of Chemical Physics} \textbf{\bibinfo{volume}{98}}, \bibinfo{pages}{8718} (\bibinfo{year}{1993}).

\bibitem[{\citenamefont{Bartlett et~al.}(1990)\citenamefont{Bartlett, Watts, Kucharski, and Noga}}]{bartlett1990non}
\bibinfo{author}{\bibfnamefont{R.~J.} \bibnamefont{Bartlett}}, \bibinfo{author}{\bibfnamefont{J.}~\bibnamefont{Watts}}, \bibinfo{author}{\bibfnamefont{S.}~\bibnamefont{Kucharski}}, \bibnamefont{and} \bibinfo{author}{\bibfnamefont{J.}~\bibnamefont{Noga}}, \bibinfo{journal}{Chemical Physics Letters} \textbf{\bibinfo{volume}{165}}, \bibinfo{pages}{513} (\bibinfo{year}{1990}).

\bibitem[{\citenamefont{Stanton}(1997)}]{stanton1997ccsd}
\bibinfo{author}{\bibfnamefont{J.~F.} \bibnamefont{Stanton}}, \bibinfo{journal}{Chemical Physics Letters} \textbf{\bibinfo{volume}{281}}, \bibinfo{pages}{130} (\bibinfo{year}{1997}).

\bibitem[{\citenamefont{L{\"o}wdin}(1962)}]{lowdin1962studies}
\bibinfo{author}{\bibfnamefont{P.-O.} \bibnamefont{L{\"o}wdin}}, \bibinfo{journal}{Journal of Mathematical Physics} \textbf{\bibinfo{volume}{3}}, \bibinfo{pages}{969} (\bibinfo{year}{1962}).

\bibitem[{\citenamefont{Kucharski and Bartlett}(1998{\natexlab{c}})}]{kucharski1998noniterative}
\bibinfo{author}{\bibfnamefont{S.~A.} \bibnamefont{Kucharski}} \bibnamefont{and} \bibinfo{author}{\bibfnamefont{R.~J.} \bibnamefont{Bartlett}}, \bibinfo{journal}{The Journal of Chemical Physics} \textbf{\bibinfo{volume}{108}}, \bibinfo{pages}{5243} (\bibinfo{year}{1998}{\natexlab{c}}).

\bibitem[{\citenamefont{McCaskey et~al.}(2020)\citenamefont{McCaskey, Lyakh, Dumitrescu, Powers, and Humble}}]{mccaskey2020xacc}
\bibinfo{author}{\bibfnamefont{A.~J.} \bibnamefont{McCaskey}}, \bibinfo{author}{\bibfnamefont{D.~I.} \bibnamefont{Lyakh}}, \bibinfo{author}{\bibfnamefont{E.~F.} \bibnamefont{Dumitrescu}}, \bibinfo{author}{\bibfnamefont{S.~S.} \bibnamefont{Powers}}, \bibnamefont{and} \bibinfo{author}{\bibfnamefont{T.~S.} \bibnamefont{Humble}}, \bibinfo{journal}{Quantum Science and Technology} \textbf{\bibinfo{volume}{5}}, \bibinfo{pages}{024002} (\bibinfo{year}{2020}).

\bibitem[{\citenamefont{Claudino et~al.}(2022)\citenamefont{Claudino, McCaskey, and Lyakh}}]{xacc_chem}
\bibinfo{author}{\bibfnamefont{D.}~\bibnamefont{Claudino}}, \bibinfo{author}{\bibfnamefont{A.~J.} \bibnamefont{McCaskey}}, \bibnamefont{and} \bibinfo{author}{\bibfnamefont{D.~I.} \bibnamefont{Lyakh}}, \bibinfo{journal}{ACM Transactions in Quantum Computing} \textbf{\bibinfo{volume}{4}}, \bibinfo{pages}{1–20} (\bibinfo{year}{2022}), \urlprefix\url{https://doi.org/10.1145/3523285}.

\bibitem[{\citenamefont{Sun et~al.}(2018)\citenamefont{Sun, Berkelbach, Blunt, Booth, Guo, Li, Liu, McClain, Sayfutyarova, Sharma et~al.}}]{sun2018pyscf}
\bibinfo{author}{\bibfnamefont{Q.}~\bibnamefont{Sun}}, \bibinfo{author}{\bibfnamefont{T.~C.} \bibnamefont{Berkelbach}}, \bibinfo{author}{\bibfnamefont{N.~S.} \bibnamefont{Blunt}}, \bibinfo{author}{\bibfnamefont{G.~H.} \bibnamefont{Booth}}, \bibinfo{author}{\bibfnamefont{S.}~\bibnamefont{Guo}}, \bibinfo{author}{\bibfnamefont{Z.}~\bibnamefont{Li}}, \bibinfo{author}{\bibfnamefont{J.}~\bibnamefont{Liu}}, \bibinfo{author}{\bibfnamefont{J.~D.} \bibnamefont{McClain}}, \bibinfo{author}{\bibfnamefont{E.~R.} \bibnamefont{Sayfutyarova}}, \bibinfo{author}{\bibfnamefont{S.}~\bibnamefont{Sharma}}, \bibnamefont{et~al.}, \bibinfo{journal}{Wiley Interdisciplinary Reviews Computational Molecular Science} \textbf{\bibinfo{volume}{8}}, \bibinfo{pages}{e1340} (\bibinfo{year}{2018}).

\bibitem[{\citenamefont{Hehre et~al.}(1969)\citenamefont{Hehre, Stewart, and Pople}}]{Hehre1969}
\bibinfo{author}{\bibfnamefont{W.~J.} \bibnamefont{Hehre}}, \bibinfo{author}{\bibfnamefont{R.~F.} \bibnamefont{Stewart}}, \bibnamefont{and} \bibinfo{author}{\bibfnamefont{J.~A.} \bibnamefont{Pople}}, \bibinfo{journal}{The Journal of Chemical Physics} \textbf{\bibinfo{volume}{51}}, \bibinfo{pages}{2657} (\bibinfo{year}{1969}).

\bibitem[{\citenamefont{Feller}(1996)}]{Feller1996}
\bibinfo{author}{\bibfnamefont{D.}~\bibnamefont{Feller}}, \bibinfo{journal}{J. Comput. Chem.} \textbf{\bibinfo{volume}{17}}, \bibinfo{pages}{1571} (\bibinfo{year}{1996}).

\bibitem[{\citenamefont{Pritchard et~al.}(2019)\citenamefont{Pritchard, Altarawy, Didier, Gibson, and Windus}}]{Pritchard2019}
\bibinfo{author}{\bibfnamefont{B.~P.} \bibnamefont{Pritchard}}, \bibinfo{author}{\bibfnamefont{D.}~\bibnamefont{Altarawy}}, \bibinfo{author}{\bibfnamefont{B.}~\bibnamefont{Didier}}, \bibinfo{author}{\bibfnamefont{T.~D.} \bibnamefont{Gibson}}, \bibnamefont{and} \bibinfo{author}{\bibfnamefont{T.~L.} \bibnamefont{Windus}}, \bibinfo{journal}{Journal of Chemical Information and Modeling} \textbf{\bibinfo{volume}{59}}, \bibinfo{pages}{4814} (\bibinfo{year}{2019}).

\bibitem[{UT2()}]{UT2}
\emph{\bibinfo{title}{Ut2: A python-based suite of coupled-cluster methods designed to converge towards an "ultimate" t2 method}}, \bibinfo{howpublished}{\url{https://github.com/zww-4855/ut2}}.

\bibitem[{\citenamefont{K{\"o}hn and Olsen}(2022)}]{kohn2022capabilities}
\bibinfo{author}{\bibfnamefont{A.}~\bibnamefont{K{\"o}hn}} \bibnamefont{and} \bibinfo{author}{\bibfnamefont{J.}~\bibnamefont{Olsen}}, \bibinfo{journal}{The Journal of Chemical Physics} \textbf{\bibinfo{volume}{157}} (\bibinfo{year}{2022}).

\bibitem[{\citenamefont{Lee et~al.}(2019)\citenamefont{Lee, Bertels, Small, and Head-Gordon}}]{lee2019kohn}
\bibinfo{author}{\bibfnamefont{J.}~\bibnamefont{Lee}}, \bibinfo{author}{\bibfnamefont{L.~W.} \bibnamefont{Bertels}}, \bibinfo{author}{\bibfnamefont{D.~W.} \bibnamefont{Small}}, \bibnamefont{and} \bibinfo{author}{\bibfnamefont{M.}~\bibnamefont{Head-Gordon}}, \bibinfo{journal}{Physical Review Letters} \textbf{\bibinfo{volume}{123}}, \bibinfo{pages}{113001} (\bibinfo{year}{2019}).

\bibitem[{\citenamefont{Windom et~al.}(2022)\citenamefont{Windom, Perera, and Bartlett}}]{windom2022benchmarking}
\bibinfo{author}{\bibfnamefont{Z.~W.} \bibnamefont{Windom}}, \bibinfo{author}{\bibfnamefont{A.}~\bibnamefont{Perera}}, \bibnamefont{and} \bibinfo{author}{\bibfnamefont{R.~J.} \bibnamefont{Bartlett}}, \bibinfo{journal}{The Journal of Chemical Physics} \textbf{\bibinfo{volume}{156}} (\bibinfo{year}{2022}).

\bibitem[{\citenamefont{Peruzzo et~al.}(2014)\citenamefont{Peruzzo, McClean, Shadbolt, Yung, Zhou, Love, Aspuru-Guzik, and O'Brien}}]{Peruzzo2014}
\bibinfo{author}{\bibfnamefont{A.}~\bibnamefont{Peruzzo}}, \bibinfo{author}{\bibfnamefont{J.}~\bibnamefont{McClean}}, \bibinfo{author}{\bibfnamefont{P.}~\bibnamefont{Shadbolt}}, \bibinfo{author}{\bibfnamefont{M.-H.} \bibnamefont{Yung}}, \bibinfo{author}{\bibfnamefont{X.-Q.} \bibnamefont{Zhou}}, \bibinfo{author}{\bibfnamefont{P.~J.} \bibnamefont{Love}}, \bibinfo{author}{\bibfnamefont{A.}~\bibnamefont{Aspuru-Guzik}}, \bibnamefont{and} \bibinfo{author}{\bibfnamefont{J.~L.} \bibnamefont{O'Brien}}, \bibinfo{journal}{Nature Communications} \textbf{\bibinfo{volume}{5}} (\bibinfo{year}{2014}).

\bibitem[{\citenamefont{Grimsley et~al.}(2019{\natexlab{b}})\citenamefont{Grimsley, Claudino, Economou, Barnes, and Mayhall}}]{grimsley2019trotterized}
\bibinfo{author}{\bibfnamefont{H.~R.} \bibnamefont{Grimsley}}, \bibinfo{author}{\bibfnamefont{D.}~\bibnamefont{Claudino}}, \bibinfo{author}{\bibfnamefont{S.~E.} \bibnamefont{Economou}}, \bibinfo{author}{\bibfnamefont{E.}~\bibnamefont{Barnes}}, \bibnamefont{and} \bibinfo{author}{\bibfnamefont{N.~J.} \bibnamefont{Mayhall}}, \bibinfo{journal}{Journal of Chemical Theory and Computation} \textbf{\bibinfo{volume}{16}}, \bibinfo{pages}{1} (\bibinfo{year}{2019}{\natexlab{b}}).

\bibitem[{\citenamefont{Casanova}(2012)}]{casanova2012avoided}
\bibinfo{author}{\bibfnamefont{D.}~\bibnamefont{Casanova}}, \bibinfo{journal}{The Journal of Chemical Physics} \textbf{\bibinfo{volume}{137}} (\bibinfo{year}{2012}).

\bibitem[{\citenamefont{Harbison}(2002)}]{harbison2002electric}
\bibinfo{author}{\bibfnamefont{G.~S.} \bibnamefont{Harbison}}, \bibinfo{journal}{Journal of the American Chemical Society} \textbf{\bibinfo{volume}{124}}, \bibinfo{pages}{366} (\bibinfo{year}{2002}).

\bibitem[{\citenamefont{Kardahakis et~al.}(2005)\citenamefont{Kardahakis, Pittner, {\v{C}}{\'a}rsky, and Mavridis}}]{kardahakis2005multireference}
\bibinfo{author}{\bibfnamefont{S.}~\bibnamefont{Kardahakis}}, \bibinfo{author}{\bibfnamefont{J.}~\bibnamefont{Pittner}}, \bibinfo{author}{\bibfnamefont{P.}~\bibnamefont{{\v{C}}{\'a}rsky}}, \bibnamefont{and} \bibinfo{author}{\bibfnamefont{A.}~\bibnamefont{Mavridis}}, \bibinfo{journal}{International Journal of Quantum Chemistry} \textbf{\bibinfo{volume}{104}}, \bibinfo{pages}{458} (\bibinfo{year}{2005}).

\bibitem[{\citenamefont{Su et~al.}(2008)\citenamefont{Su, Wu, Shaik, and Hiberty}}]{su2008valence}
\bibinfo{author}{\bibfnamefont{P.}~\bibnamefont{Su}}, \bibinfo{author}{\bibfnamefont{W.}~\bibnamefont{Wu}}, \bibinfo{author}{\bibfnamefont{S.}~\bibnamefont{Shaik}}, \bibnamefont{and} \bibinfo{author}{\bibfnamefont{P.~C.} \bibnamefont{Hiberty}}, \bibinfo{journal}{ChemPhysChem} \textbf{\bibinfo{volume}{9}}, \bibinfo{pages}{1442} (\bibinfo{year}{2008}).

\bibitem[{\citenamefont{Peterson and Woods}(1989)}]{peterson1989ground}
\bibinfo{author}{\bibfnamefont{K.~A.} \bibnamefont{Peterson}} \bibnamefont{and} \bibinfo{author}{\bibfnamefont{R.~C.} \bibnamefont{Woods}}, \bibinfo{journal}{The Journal of Chemical Physics} \textbf{\bibinfo{volume}{90}}, \bibinfo{pages}{7239} (\bibinfo{year}{1989}).

\bibitem[{\citenamefont{Zhai et~al.}(2007)\citenamefont{Zhai, Wang, Li, and Wang}}]{zhai2007vibrationally}
\bibinfo{author}{\bibfnamefont{H.-J.} \bibnamefont{Zhai}}, \bibinfo{author}{\bibfnamefont{L.-M.} \bibnamefont{Wang}}, \bibinfo{author}{\bibfnamefont{S.-D.} \bibnamefont{Li}}, \bibnamefont{and} \bibinfo{author}{\bibfnamefont{L.-S.} \bibnamefont{Wang}}, \bibinfo{journal}{The Journal of Physical Chemistry A} \textbf{\bibinfo{volume}{111}}, \bibinfo{pages}{1030} (\bibinfo{year}{2007}).

\end{thebibliography}

\end{document}


\title{Supplementary Material for Toward the ``platinum standard" of quantum chemistry on quantum computers: perturbative quadruple corrections in unitary coupled cluster theory}
\author{Zachary W. Windom$^{1, 2}$, Luke Bertels$^{2}$, Daniel Claudino$^2$\footnote{\href{mailto:claudinodc@ornl.gov}{claudinodc@ornl.gov}}, and Rodney J. Bartlett$^1$}
\affiliation{$^1$Quantum Theory Project, University of Florida, Gainesville, FL, 32611, USA \\
$^2$Computational Sciences and Engineering Division,\ Oak\ Ridge\ National\ Laboratory,\ Oak\ Ridge,\ TN,\ 37831,\ USA}
\maketitle

\section{Trotter operator ordering}

Recognizing that distinct operator orderings of the tUCC ansatz necessarily define distinct wavefunctions, prior work suggested that the set of so-called ``disentangled" ansatz formed from each distinct operator ordering leads to a family of ansatz that do not necessarily coincide with the full operator, UCC wavefunction and/or energy.\cite{evangelista2019exact} Subsequent work explored the numerical impact of operator with respect to the corresponding trotterized UCCSD energy, and found that distinct operator orderings lead to significant energetic differences - particularly in ``strongly correlated" regimes - that can frequently exceed 1 mH.\cite{grimsley2019trotterized} Although the protocol suggested by this work advocates for an ordering where amplitudes modulating HOMO-LUMO-like excitations should take precedence and act first on the reference determinant, the current work highlights that such operator ordering strategies can lead to a - somewhat premature - ``variational catastrophe" of the perturbative [Q-6] correction. To be clear, Ref. \cite{grimsley2019trotterized}  suggests that the ``best" operator ordering defining tUCCSD was one in which the (presumably) ``most important" amplitudes were considered first, meaning $exp(\tau_2)$ acts on $\ket{0}$ before $exp(\tau_1)$ where the individual amplitudes $t_i^a, t_{ij}^{ab}$ that modulate excitations between energetically close, adjacent occupied/unoccupied orbitals act on $\ket{0}$ first.

However, it was previously found that this particular way of organizing the tUCC ansatz can lead to ill-behaved perturbative corrections in static correlation regimes. Ref. \citenum{windom2024new} briefly considered the impact operator ordering has on the ensuing perturbative correction, and showed the behavior of the PT-based correction can be significantly improved if any other operator ordering is adopted. To this end, the current work initially examines the operator ordering recommended in Ref. \citenum{grimsley2019trotterized} to define the tUCCSDT ansatz. This  proved entirely adequate in scans of the LiF and NF PES.

\begin{figure}[ht]
 \centering
 \includegraphics[width=\columnwidth]{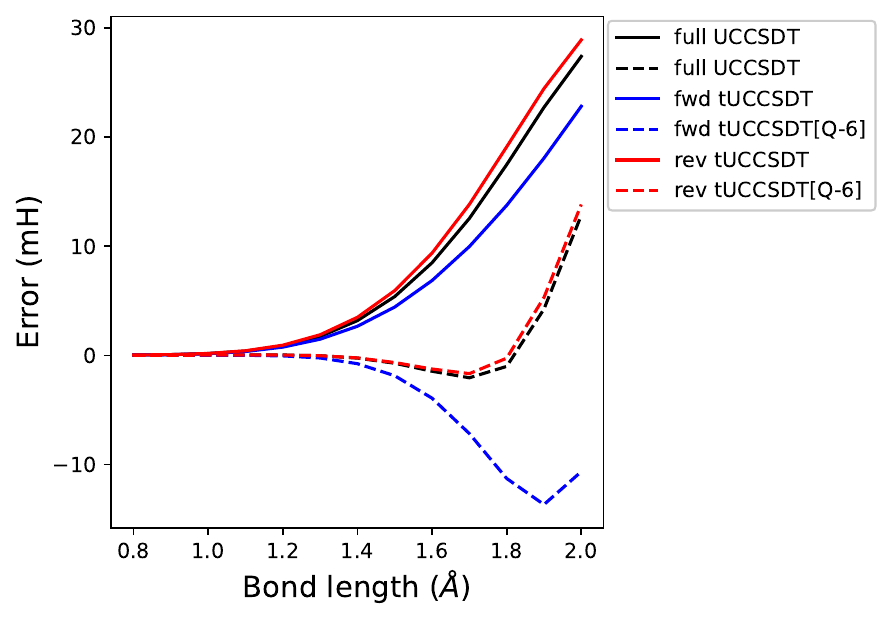}
 \caption{Comparison of method error across the BO- PES with respect to FCI (mH). We show the full operator (full), as well as two trotterized ansatz that are differentiate by their operator ordering; eg ``forward" (fwd) and ``reverse" (rev)  {\color{black}Errors are reported with respect to FCI.}}
 \label{fig:bominuserrors}
\end{figure}

\begin{figure}[ht]
 \centering
 \includegraphics[width=\columnwidth]{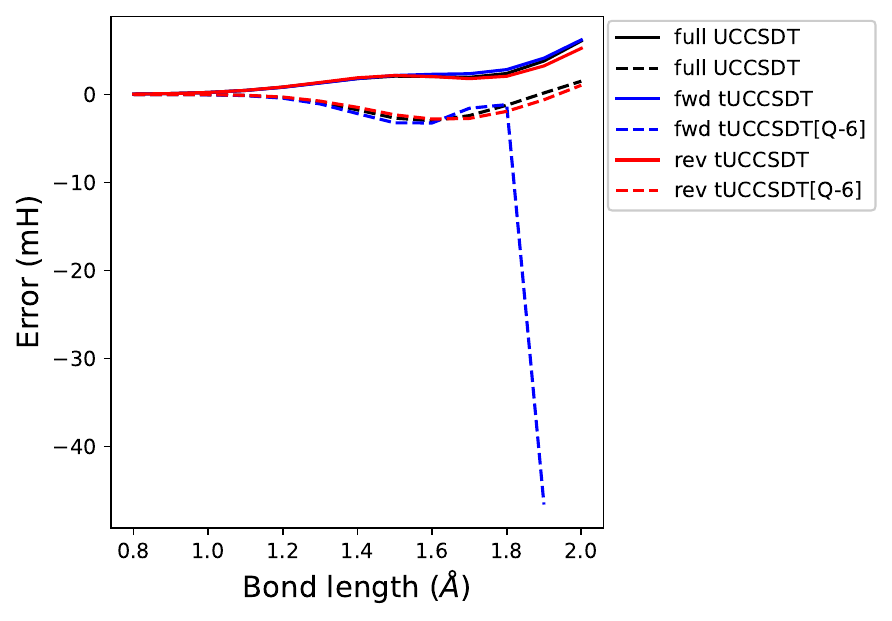}
 \caption{Comparison of method error across the N$_2$ PES with respect to FCI (mH). We show the full operator (full), as well as two trotterized ansatz that are differentiate by their operator ordering; eg ``forward" (fwd) and ``reverse" (rev)  {\color{black}Errors are reported with respect to FCI.}}
 \label{fig:N2errors}
\end{figure}

For the remaining examples, however, we found that this particular operator ordering choice leads to spurious [Q-6] corrections in the stretched regime. Consequently, we explored another operator ordering that  reverses  the ordering advocated by Ref. \cite{grimsley2019trotterized}, wherein $\tau_1$ is optimized before $\tau_2$ which is optimized before $\tau_3$. Figures \ref{fig:bominuserrors}, \ref{fig:N2errors}, and \ref{fig:N2errors} compares the error of [Q-6]-corrected ansatz with respect to FCI, where the perturbative correction is constructed with respect to the full UCCSDT (full) operator as well as trotterized UCCSDT ansatz that have adopted the default operator ordering proposed by Ref. \cite{grimsley2019trotterized}(fwd) as well as the reverse of this default ordering (rev).  For these examples, we found that the default ordering leads to either spurious results and/or non-variational catastrophes of the [Q-6] correction. The reverse of this operator ordering, however, leads to [Q-6] corrections that are more comparable to the corresponding, full UCCSDT operator results. Consequently, the primary text studies the tUCCSDT ansatz defined by the reversed operator ordering scheme for the scans of the BO-, N$_2$, and O$_2$ PES.

\begin{figure}[ht]
 \centering
 \includegraphics[width=\columnwidth]{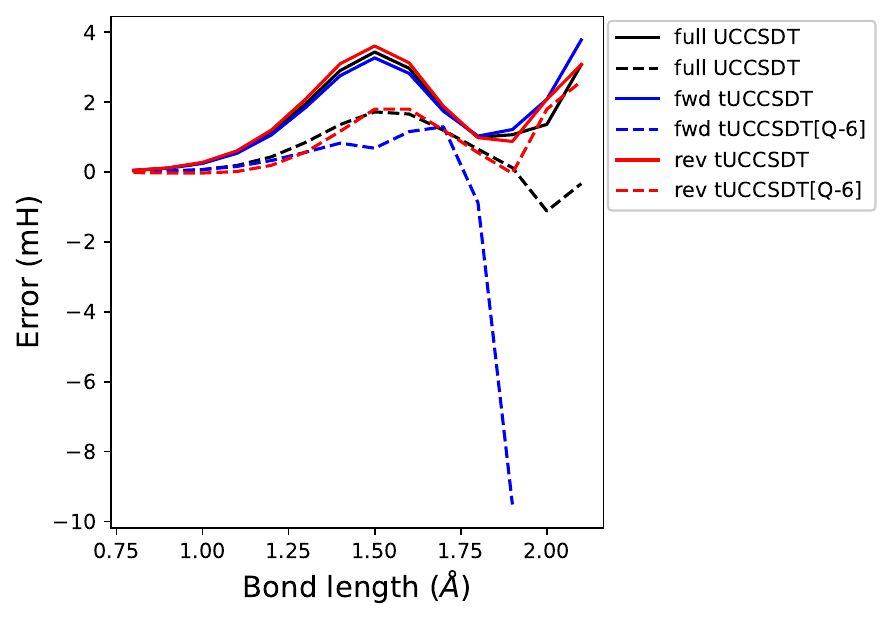}
 \caption{Comparison of method error across the O$_2$ PES with respect to FCI (mH). We show the full operator (full), as well as two trotterized ansatz that are differentiate by their operator ordering; eg ``forward" (fwd) and ``reverse" (rev)  {\color{black}Errors are reported with respect to FCI.}}
 \label{fig:O2errors}
\end{figure}

\thanks{This manuscript has been authored by UT-Battelle, LLC, under Contract No.~DE-AC0500OR22725 with the U.S.~Department of Energy. The United States Government retains and the publisher, by accepting the article for publication, acknowledges that the United States Government retains a non-exclusive, paid-up, irrevocable, world-wide license to publish or reproduce the published form of this manuscript, or allow others to do so, for the United States Government purposes. The Department of Energy will provide public access to these results of federally sponsored research in accordance with the DOE Public Access Plan.}



\bibliography{biblio}